\documentclass[12pt]{article}
\pdfoutput=1
\usepackage{amsmath}
\usepackage{amssymb}
\usepackage{graphicx}
\usepackage{subcaption}
\usepackage{url}
\usepackage{arydshln}
\usepackage[sort&compress, numbers, merge]{natbib}

\newcommand{\be}{\begin{equation}}
\newcommand{\ee}{\end{equation}}

\newcommand{\bea}{\begin{eqnarray}}
\newcommand{\eea}{\end{eqnarray}}
\newcommand{\mat}{\begin{pmatrix}}
\newcommand{\rix}{\end{pmatrix}}

\renewcommand{\bar}{\overline}

\def\beq{\begin{equation}}
\def\eeq{\end{equation}}
\def\beqn{\begin{equation}}
\def\eeqn{\end{equation}}

\usepackage[colorlinks=true, linkcolor=magenta, citecolor=blue,
urlcolor=black]{hyperref}

\begin{document}

\begin{titlepage}
\begin{center}

\vspace{3.0cm}
{\large \bf Modification of the Sommerfeld effect due to coannihilator decays}

\vspace{1.0cm}
{\bf Feng Luo} 
\vspace{1.0cm}

{\it
{School of Physics and Astronomy, Sun Yat-sen University, Zhuhai 519082, China}\\
}

\vspace{1cm}
\abstract{
In dark matter coannihilation scenarios, the Sommerfeld effect of coannihilators is usually important in calculations of dark matter thermal relic abundance. 
Due to decays of coannihilators, two annihilating coannihilators can only approach each other from a finite initial separation. 
While conventional derivations of Sommerfeld factors essentially assume an infinite initial separation, Sommerfeld factors for a pair of annihilating coannihilators may be different. 
We find that the Sommerfeld factor for a pair of coannihilators is less prominent than the one obtained without considering decays. 
The modification of the Sommerfeld factor may result in a change of the dark matter thermal relic abundance well beyond the percent level. 
}
\end{center}

\end{titlepage}

\setcounter{footnote}{0}

\section{Introduction}

The cold dark matter (DM) density of the Universe has been determined by observation to a percent level accuracy, $\Omega_c h^2 = 0.120\pm0.001$~\cite{Planck:2018vyg}.
It is the only precise quantity we know about DM, and certainly it deserves thorough investigations. 
One way to obtain this value in theory is through the thermal freeze-out mechanism~\cite{Lee:1977ua,Kolb:1990vq}. 
This mechanism is attractive, because it has successfully helped to explain two other important relics of our Universe --- the cosmic microwave background and the light elements from Big-Bang nucleosynthesis.  

During freeze-out, typical velocities of cold DM particles are non-relativistic. 
If there is some long-range interaction between DM particles, the two-body wave function of an annihilating DM pair is modified from a plane wave, and therefore the annihilation cross section and consequently the DM thermal relic abundance are affected. 
This is the Sommerfeld effect~\cite{sommerfeld1931, hep-ph/0610249}. 
Roughly speaking, the annihilation cross section is the product of the so called Sommerfeld factor and the bare annihilation cross section. The latter is calculated through the usual relativistic quantum field theory approach by using plane wave functions. 
The Sommerfeld factor can be obtained in non-relativistic quantum mechanics framework by solving for the scattering wave function of a particle moving in a long-range potential. 
It is bigger than 1 (i.e., Sommerfeld enhancement) if the potential is attractive, while it is smaller than 1 (i.e., Sommerfeld suppression) if the potential is repulsive. 
A stronger attractive or repulsive potential gives a more significant Sommerfeld enhancement or suppression, because it results in a larger modification of the scattering wave function relative to the plane wave function. 
A small relative velocity $v_{rel}$ between the two annihilating particles is another indispensable ingredient in getting a significant Sommerfeld enhancement or suppression. 
This can be understood intuitively. 
A smaller $v_{rel}$ means that the potential can act on the annihilating particles for a longer time before they meet and annihilate, so that a larger accumulation of the modification of the two-body wave function can be achieved. 
For example, at small $v_{rel}$, the $s$-wave Sommerfeld factor for a Coulomb-like potential $V(r) = - \frac{\alpha}{r}$ of an infinite range is approximately $\frac{2 \pi \alpha}{v_{rel}}$ for an attractive case ($\alpha > 0$), and it is about $-\frac{2 \pi \alpha}{v_{rel}} e^{\frac{2 \pi \alpha}{v_{rel}}}$ for a repulsive one ($\alpha < 0$). 
The Sommerfeld effect on DM thermal relic abundance calculations has been well-studied (see e.g.~\cite{Cirelli:2007xd, Hryczuk:2010zi}). 
Also, Sommerfeld factors may be crucial in explaining DM indirect detection results (see e.g.~\cite{Arkani-Hamed:2008hhe, Lattanzi:2008qa}). 

Another important add-on during freeze-out is coannihilation~\cite{Griest:1990kh}. 
Some particle species (i.e., coannihilators), which are slightly heavier than the DM particles, freeze out together with the latter.
Through scatterings and (inverse-)decays, coannihilators and DM particles can interconvert to each other. Consequently, the ratio of their number densities equals the equilibrium value at temperatures during freeze-out, if the interconversion rate is larger than the Hubble expansion rate. 
Depending on the relative sizes of the DM-DM, DM-coannihilator and coannihilator-coannihilator (co)annihilation cross sections, the mass difference between the coannihilator and the DM particle, and their degrees of freedom, the DM relic abundance can be smaller or larger than if no coannihilator species exist~\cite{Profumo:2006bx}.
Coannihilation is a feasible and sometimes unavoidable feature in many theories beyond the Standard Model (BSM), including scenarios in supersymmetry~\cite{Ellis:1998kh, Beneke:2016ync} and Universal Extra Dimensions~\cite{Servant:2002aq, Burnell:2005hm}. 
Various coannihilation models in DM searches have also been extensively studied (see e.g.~\cite{Baker:2015qna}). 

If there is some long-range interaction between coannihilators, the Sommerfeld effect of annihilating coannihilators needs to be taken into account in the calculation of DM thermal relic abundance. 
In fact, compared to DM particles, it is more often to have long-range interactions between coannihilators. 
For example, coannihilators can be electrically and/or color charged~\cite{1503.07142, 1510.03498, Ellis:2018jyl, 1812.02066}, but usually the DM cannot be.
Long-range interactions felt by dark sector particles can arise not only from Standard Model gauge and Higgs interactions, but also from the exchanges of new gauge bosons or scalars in BSM models~\cite{Feng:2010zp, Cirelli:2016rnw}. 

Sommerfeld factors for coannihilators are conventionally computed using the same formula as for DM particles.
However, we pointed out in a previous work~\cite{Cui:2020ppc} that due to interconversions between coannihilators and DM particles, Sommerfeld factors are closer to 1 than if the impact of interconversions are not taken into account. 
The reason is the following. 
To have the two-body wave function of a coannihilator pair significantly modified from a plane wave, the two particles need to approach each other from an initial separation large enough compared to the characteristic distance scale of the long-range interaction, which is given by the inverse of the relative momentum $(\mu v_{rel})^{-1}$, where $\mu$ is the reduced mass of the two particles. 
Suppose the interconversion rate is $\Gamma_{con}$, then the typical initial separation is $v_{rel}/(2 \Gamma_{con})$~\footnote{That $v_{rel}/(2 \Gamma_{con})$ being the typical initial separation of two annihilating coannihilators may be easier to be understood by considering the reverse process, in which two coannihilator particles are produced in the final state. In the reverse process, $v_{rel}/(2 \Gamma_{con})$ is the typical separation that the two particles can achieve in their center of mass frame, before one of them converts into a DM particle.}. 
Therefore, when $v_{rel} < \sqrt{2 \Gamma_{con}/\mu}$ the Sommerfeld effect is less effective~\footnote{The same result can be obtained by an equivalent explanation used in discussing the effect of long-range force between unstable heavy charged particles produced in colliders~\cite{hep-ph/9501214, Fadin:1993kg, Bardin:1993mc, Fadin:1993kt}. 
For a pair of heavy charged particles, the time scale for the Coulomb-like force to have an impact on the production cross section is given by the inverse of the relative kinetic energy $(\sim \mu v_{rel}^2 /2)^{-1}$. If it is longer than the lifetime of the particles, the impact is reduced.}. 
In~\cite{Cui:2020ppc} we took $\sqrt{2 \Gamma_{con}/\mu}$ as a cut-off velocity, below which the Sommerfeld factor was switched off (that is, set it to be 1). 
This approach captures the key physics, but certainly it can be improved.

In this paper, we introduce another method to investigate the coannihilator-DM interconversion effect on Sommerfeld factors of coannihilators, and consequently on the DM thermal relic abundance in coannihilation scenarios. 
The idea is the following. 
Due to coannihilator-DM interconversions, two annihilating coannihilators cannot approach each other from an infinite separation, otherwise they do not have a chance to meet. 
For an annihilation event to occur, the two particles have to come together from a finite initial separation, and they can feel the long-range potential produced by themselves only from their initial separation till they meet. 
For two particles with a finite initial separation, the scattering wave function is less modified from the plane wave, compared to the case when the initial separation is infinite. 
For a given annihilating coannihilator pair, the interconversion rate determines the probability distribution of the initial separation~\footnote{It may be easier to be understood by considering the reverse process, in which the interconversion rate determines the probability distribution of the separation between the two coannihilators produced in the final state.}. 
The Sommerfeld factors obtained after taking into account this distribution are referred to as rate-averaged Sommerfeld factors (RASFs) in the following. These RASFs are to be compared with the conventional Sommerfeld factors obtained without considering interconversions. We find that for the same long-range interaction strength and the same relative velocity, RASFs are less prominent (i.e., closer to 1) due to interconversions. 

The interconversion rate is the sum of the coannihilator (into DM) decay rate and coannihilator-DM scattering rate. These rates are determined by the same coupling between the coannihilator and the DM particle. The decay rate is usually larger than the scattering rate, unless for situations where the coannihilator and the DM particle are very degenerate in mass. 
Therefore, without loss of the physics we want to present, to simplify our discussion we only take into account the decay rate in this work, and we consider situations where the mass difference between the coannihilator and the DM particle is not very small. 

The rest of the paper is organized as follows.
In section~\ref{sec:classcal_mechanics_analogy}, we provide an analogy in classical mechanics to illustrate the physics. 
In section~\ref{sec:finite-range_averaged_Sommerfeld_factor}, we discuss Sommerfeld factors obtained after taking into account the finite initial separation of two annihilating particles, leaving detailed derivations in Appendix~\ref{sec:appendix A}. 
Then we use the coannihilator decay rate to get the RASFs. 
In section~\ref{sec:thermally_averaged_Sommerfeld_factor}, by further accounting for  the velocity distribution of coannihilators for a given temperature, we compute the thermally averaged Sommerfeld factor.
We apply the result to a simple coannihilation scenario, and calculate the relative changes of the DM thermal relic abundance due to the modification of Sommerfeld factors induced by decays of coannihilators. 
We summarize our conclusions in section~\ref{sec:summary}.
As a proof of concept, we focus on the $s$-wave Sommerfeld factor for a Coulomb potential in the main text, and we discuss how the results may change for a Hulth\'en potential in Appendix~\ref{sec:appendix B}. 
In Appendix~\ref{sec:appendix C}, the viability of using the quantum mechanical method to determine Sommerfeld factors in the case of annihilating particle decays is discussed.

\section{An analogy in classical mechanics}
\label{sec:classcal_mechanics_analogy}

In~\cite{Arkani-Hamed:2008hhe} a simple analogy in classical mechanics was provided to facilitate the understanding of the Sommerfeld enhancement. 
Consider a point particle coming from infinity and moving towards a star under the sole influence of gravity. 
The star has a radius $R$ and a mass $M$. 
The velocity of the particle at infinity is $v$. 
Using conservations of angular momentum and energy, one can calculate the largest impact parameter, $b_{max}$, for which the particle can hit the star, 
\beq
\Big(\frac{b_{max}}{R} \Big)^2 = 1 + \frac{2 G M}{v^2}\frac{1}{R} \,,
\eeq
where $G$ is the gravitational constant. 
If one defines the cross section as the area that the particle can hit the star, then without gravity it is $\pi R^2$, while with gravity it is $\pi b_{max}^2$. Therefore the enhancement of the cross section due to gravity is $\big(1 + \frac{2 G M}{v^2}\frac{1}{R}\big)$. 
The enhancement is larger for smaller $v$. If the interaction strength $G$ could be made larger, the enhancement would be also larger. 
Indeed, a large long-distance interaction strength and a small relative velocity are the two decisive factors to give rise to a significant Sommerfeld enhancement in quantum mechanics. 

Now, if the particle does not come from infinity, but instead it is released from a finite distance $r_0$ ($r_0>R$) with the same initial velocity $v$, then 
\beq
\Big(\frac{b_{max}}{R} \Big)^2 = 1 + \frac{2 G M}{v^2} \Big(\frac{1}{R} - \frac{1}{r_0} \Big) \,.
\eeq
Therefore finite $r_0$ results in a smaller enhancement of the cross section. 
It illustrates another important ingredient to obtain a large Sommerfeld enhancement:  
the long-range interaction needs to act on~\footnote{The ``act on'' is accounted for after the particle has been released. Although the gravity between the particle and the star is already there before the release, its effect is neutralized by the force that holds the particle.} 
the particle for a long distance.  
One can also see that $b_{max}/R \to 1$ as $r_0 \to R$. 
It indicates that there is no Sommerfeld enhancement, if the long-range interaction cannot act on the particle at all. 

The same derivation can be applied to a long-range repulsive interaction as well. 
Consider a point-like charged particle with a mass $m$ and a charge $q$ moving towards a same-sign charged ball with a radius $R$ and a charge $Q$. 
The particle is released with an initial velocity $v$ from a distance $r_0$ ($r_0>R$).
By conservations of angular momentum and energy, one  obtains the smallest impact parameter, $b_{min}$, for which the particle can miss the ball,    
\beq
\Big(\frac{b_{min}}{R} \Big)^2 = 1 + \frac{q Q}{2 \pi m v^2} \Big(\frac{1}{r_0} - \frac{1}{R} \Big) \,,
\eeq
where we have used that the Coulomb repulsive potential at a distance $r$ is $\frac{q Q}{4 \pi r}$. 
For the same $v$, a finite $r_0$ results in a larger $b_{min}$ as opposed to releasing the particle from infinity. 
Also, $b_{min}/R \to 1$ as $r_0 \to R$.
It indicates that the Sommerfeld suppression is less effective when the long-range repulsive interaction does not act on the particle for a long distance. 

We conclude that for both attractive and repulsive cases, long-range interactions have less of an impact when the particle is released from a finite distance as opposed to infinity.

\section{Sommerfeld factors modified by particle decays} 
\label{sec:finite-range_averaged_Sommerfeld_factor}

In this work, we use a semi-classical approach to study the particle decay effect on Sommerfeld factors. 
Consider two massive particles moving towards each other. 
They annihilate (or in general, collide) when they meet.  
Suppose either of them has a non-zero decay rate.  
If a decay occurs before they meet, the annihilation cannot happen.  
Therefore, in order for the annihilation to happen, the initial separation between the two particles has to be finite.  

Now let's assume that there is a long-range interaction between the two particles. 
The Sommerfeld factor for an annihilating pair can be calculated by solving for the scattering wave function of the Schr\"odinger equation. 
In standard calculations, the two particles in the scattering problem are approaching one another from infinity, and a long-range force is acting on them across an infinite distance.
Note that this force is generated by the two particles themselves. 
Consequently, due to decays this force can only act on the two particles for a finite distance, since the initial separation of the two particles has to be finite. 
Therefore, we derive the Sommerfeld factor by studying a scattering problem for a truncated long-range potential, i.e., a finite-range potential. Beyond the truncation distance $r_0$, the potential is set to zero. $r_0$ is just the initial separation of the two annihilating particles. 
The procedure to obtain the Sommerfeld factor for a generic finite-range central-force potential is detailed in Appendix~\ref{sec:appendix A}. 
To illustrate the idea, in the main text we focus on the $s$-wave Sommerfeld factor for a finite-range Coulomb potential (Eq.~(\ref{eq:Coulomb})).

Before we proceed, some remarks about this finite-range potential should be made. 
From the perspective of quantum field theory, a long-range force is generated by the exchanges of some light mediator between the two annihilating particles. 
In particular, a Coulomb-like potential is generated by the exchanges of some massless mediator. 
We note that the truncation $r_0$ we introduced does not indicate that we make any change of the light mediator. 
It is just a convenient way to capture the physics that the two annihilating particles can only feel the {\it infinite} long-range force across a {\it finite} distance. 
Also, a truncation in the potential enables us to use the standard procedure in studying the scattering problem in non-relativistic quantum mechanics. 
In this framework, after reducing the two particle scattering problem to a problem of a particle with a reduced mass being scattered by a central-force potential, the particle is assumed to come from infinity and then go to infinity. 
We put the information of the finite distance by truncating the potential~\footnote{Alternatively, one may study a scattering problem in which the particle is released from a finite distance. Instead of introducing a truncation to take into account the finite distance, one may include this information by modifying the conventional boundary condition without imposing large-distance asymptotics~\cite{Liu:2014cta}.}. 

The explicit expression of the $s$-wave Sommerfeld factor for a finite-range Coulomb potential is given in Eq.~(\ref{eq:s-wave_Coulomb}), in which $v_{rel}$, $r_0$, and the potential strength $\alpha$ appear in two combinations, $\epsilon_v \equiv \frac{v_{rel}}{\alpha}$ and $\eta_0 \equiv \alpha \mu r_0$. $\mu$ is the reduced mass of the two particles. 
$\alpha$ is greater (less) than $0$ for an attractive (a repulsive) potential. 
$|\eta_0|$ can be understood as the initial separation measured in unit of the Bohr radius $(|\alpha| \mu)^{-1}$.
In the limit $r_0 \to \infty$,
\beq
S_{0_{Coulomb}} \overset{r_0 \to \infty}{\longrightarrow} e^{\pi/\epsilon_v} \frac{\pi/\epsilon_v}{\sinh(\pi/\epsilon_v)} = \frac{2 \pi/\epsilon_v}{1 - e^{-2 \pi/\epsilon_v}} \equiv S_{r_0 \to \infty} \,.
\eeq
$S_{r_0 \to \infty}$  is the familiar result for a Coulomb potential when the force can act on the two particles across an infinite distance~\cite{Iengo:2009ni, Cassel:2009wt}.  
In the limit $r_0 \to 0$, 
\beq
S_{0_{Coulomb}} \overset{r_0 \to 0}{\longrightarrow} 1 \,,
\label{eq:r0_to_0_limit}
\eeq
as expected, since in this limit the two particles do not feel the potential at all. 

The $r_0$ in $S_{0_{Coulomb}}$ should be averaged over in order to take into account the probabilistic nature of decays. 
After a period of time $t$, the separation between the two annihilating particles changes by $R = v_{rel} t$, if they have not met yet.  
Because of decays, however, the chance that the pair still exists after time $t$ is $e^{-\Gamma R/v_{rel}}$, where $\Gamma$ is the sum of decay rates of the two particles.
Then $|d (e^{-\Gamma R/v_{rel}})| = e^{-\Gamma R/v_{rel}}\frac{\Gamma}{v_{rel}} dR$ is the probability that a decay occurs when the separation changes by an amount between $R$ and $R + dR$. 
Therefore, taking into account particle decays, the $s$-wave Sommerfeld factor for a particle pair with a given $v_{rel}$ and $\Gamma$ is  
\beq
\bar{S}_{0_{Coulomb}} = \int_0^{\infty} e^{-\Gamma \frac{r_0}{v_{rel}}} \frac{\Gamma}{v_{rel}} S_{0_{Coulomb}} dr_0 \,,
\label{eq:sbar0Coulomb}
\eeq
where the bar symbol in $\bar{S}_{0_{Coulomb}}$ indicates that the Sommerfeld factor is obtained after averaging over $r_0$. 
This equation may be easier to be understood by considering the reverse process, namely, two massive particles are produced and there is some long-range force between them until one of them decays. 
We note that we neglect the velocity dependence of $\Gamma$, that is, we neglect relativistic effect. The reason is that during and after freeze-out the typical $v_{rel}$ is non-relativistic. 
Also, this is consistent with our calculation of $S_{0_{Coulomb}}$, which is obtained in the framework of non-relativistic quantum mechanics. 

$\bar{S}_{0_{Coulomb}}$ is the $s$-wave rate-averaged Sommerfeld factors (RASF), which is to be compared with the conventional $s$-wave Sommerfeld factor obtained without considering interconversions, namely, $S_{r_0 \to \infty}$. 

\begin{figure}
\begin{center}
\begin{tabular}{c c}
\hspace{-1.0cm}
\includegraphics[height=5.8cm]{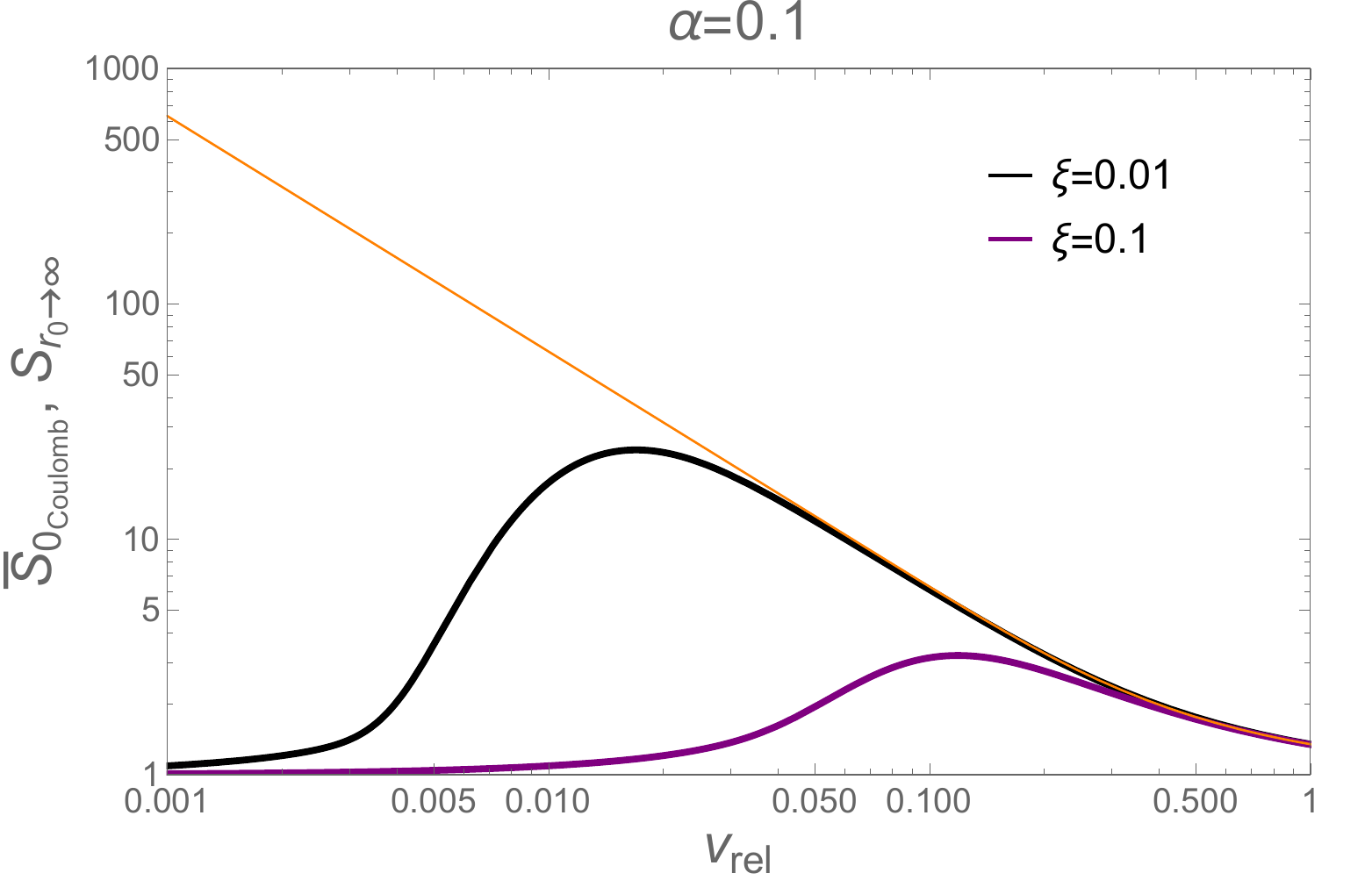} & 
\hspace{-0.2cm}
\includegraphics[height=5.8cm]{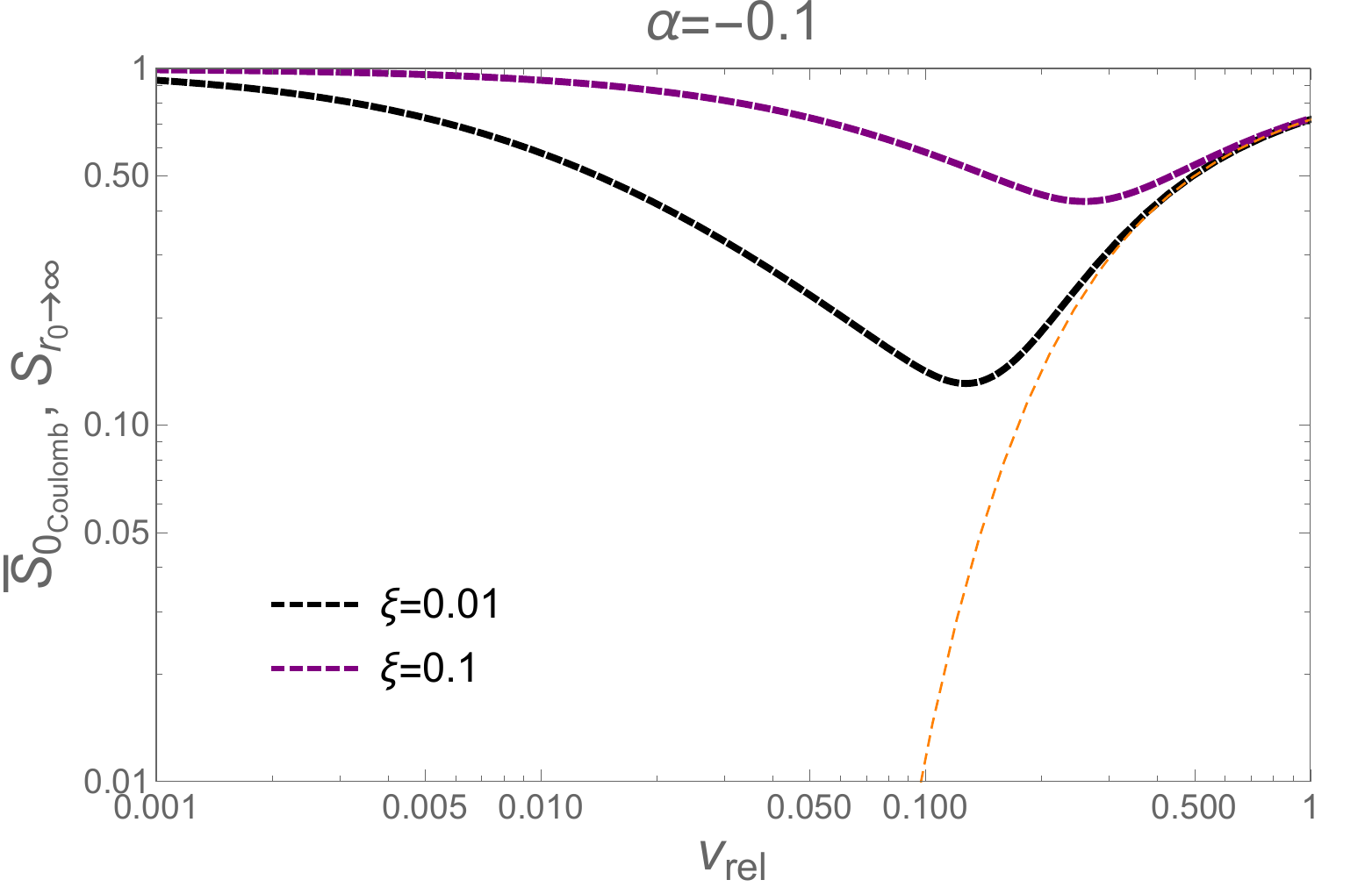} \\
\end{tabular}
\end{center}   
\begin{center}
\begin{tabular}{c c}
\hspace{-1.0cm}
\includegraphics[height=5.8cm]{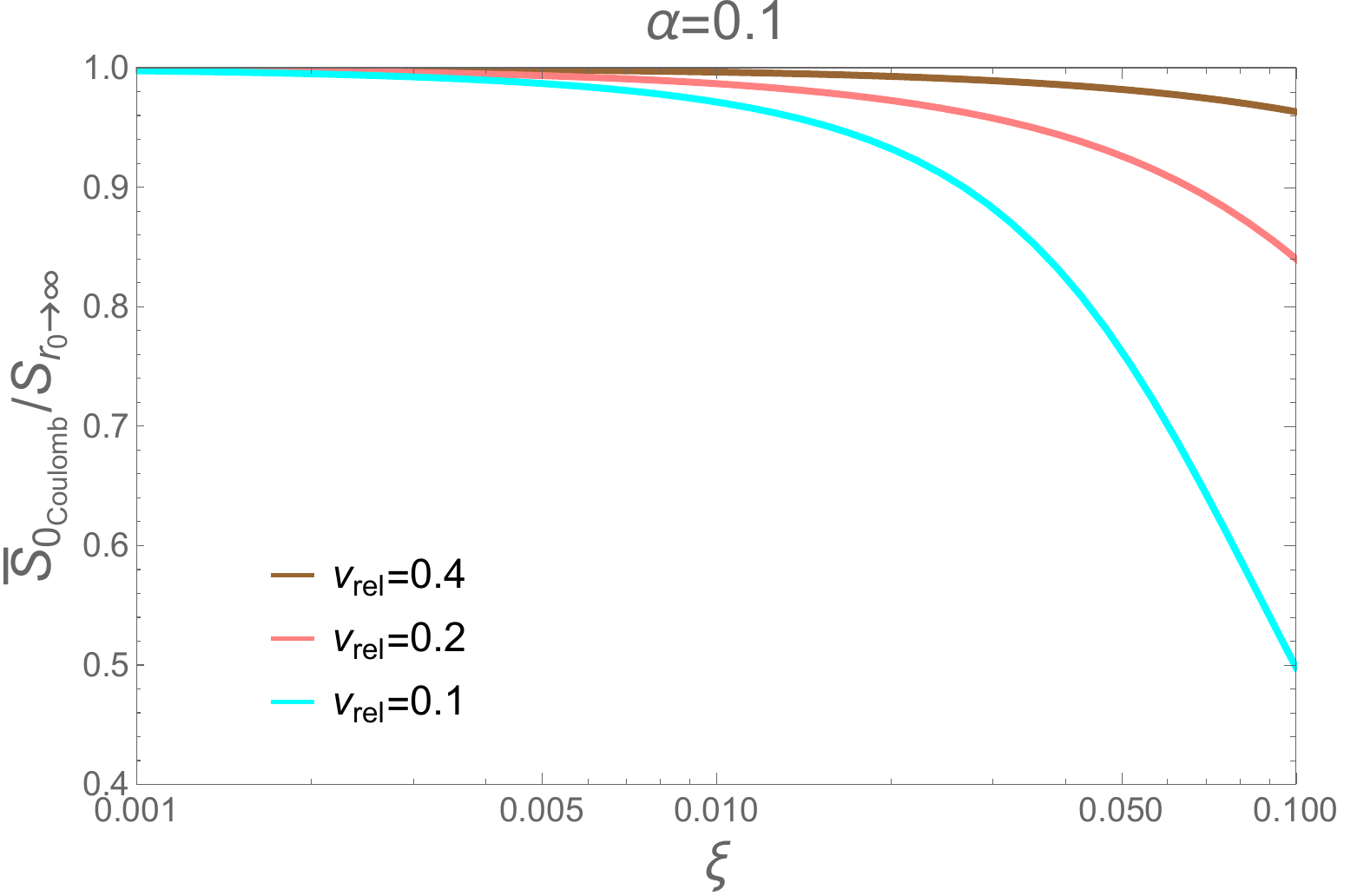} & 
\hspace{-0.2cm}
\includegraphics[height=5.8cm]{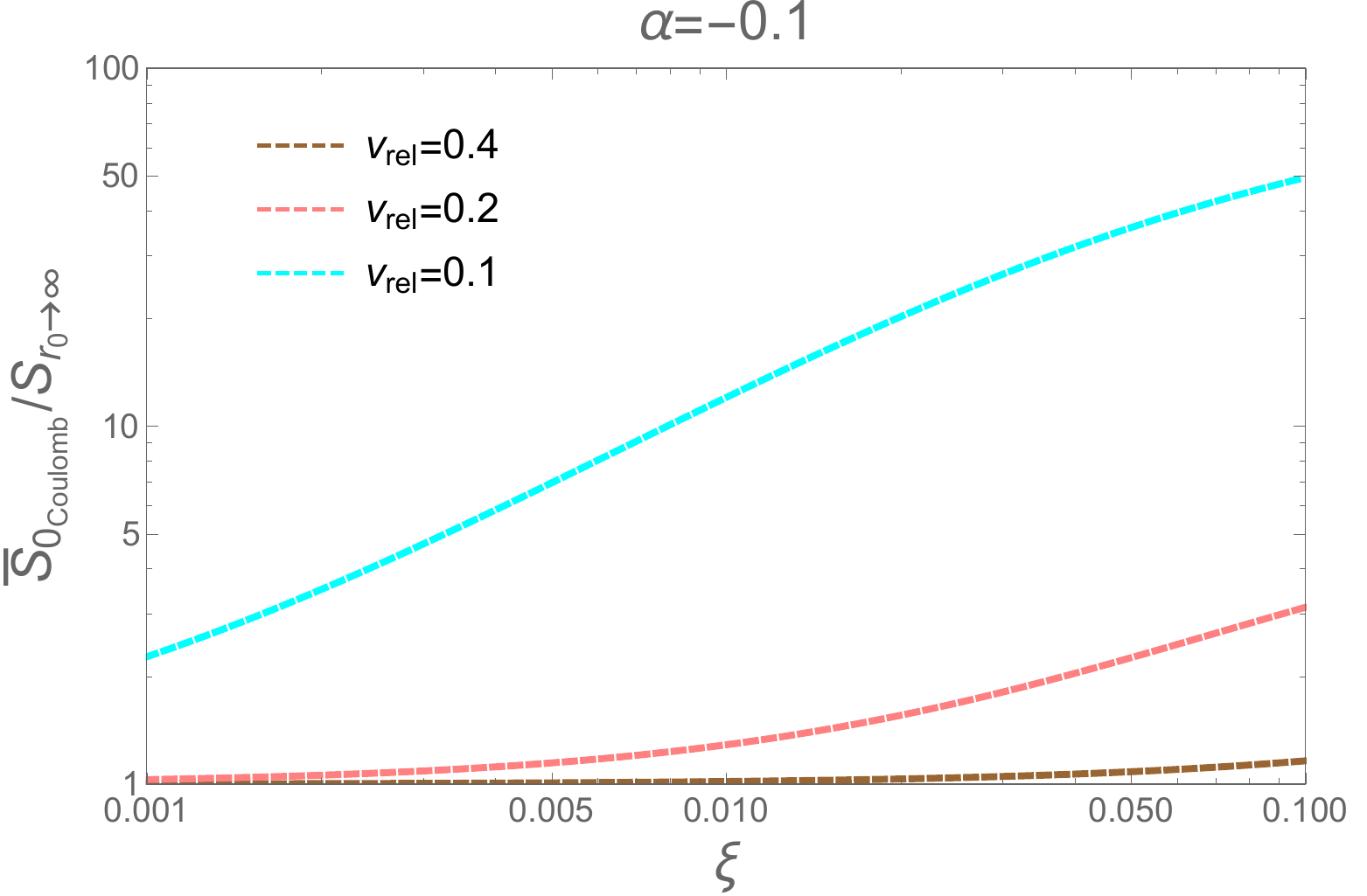} \\
\end{tabular}
\end{center}   
\begin{center}
\begin{tabular}{c c}
\hspace{-1.0cm}
\includegraphics[height=5.8cm]{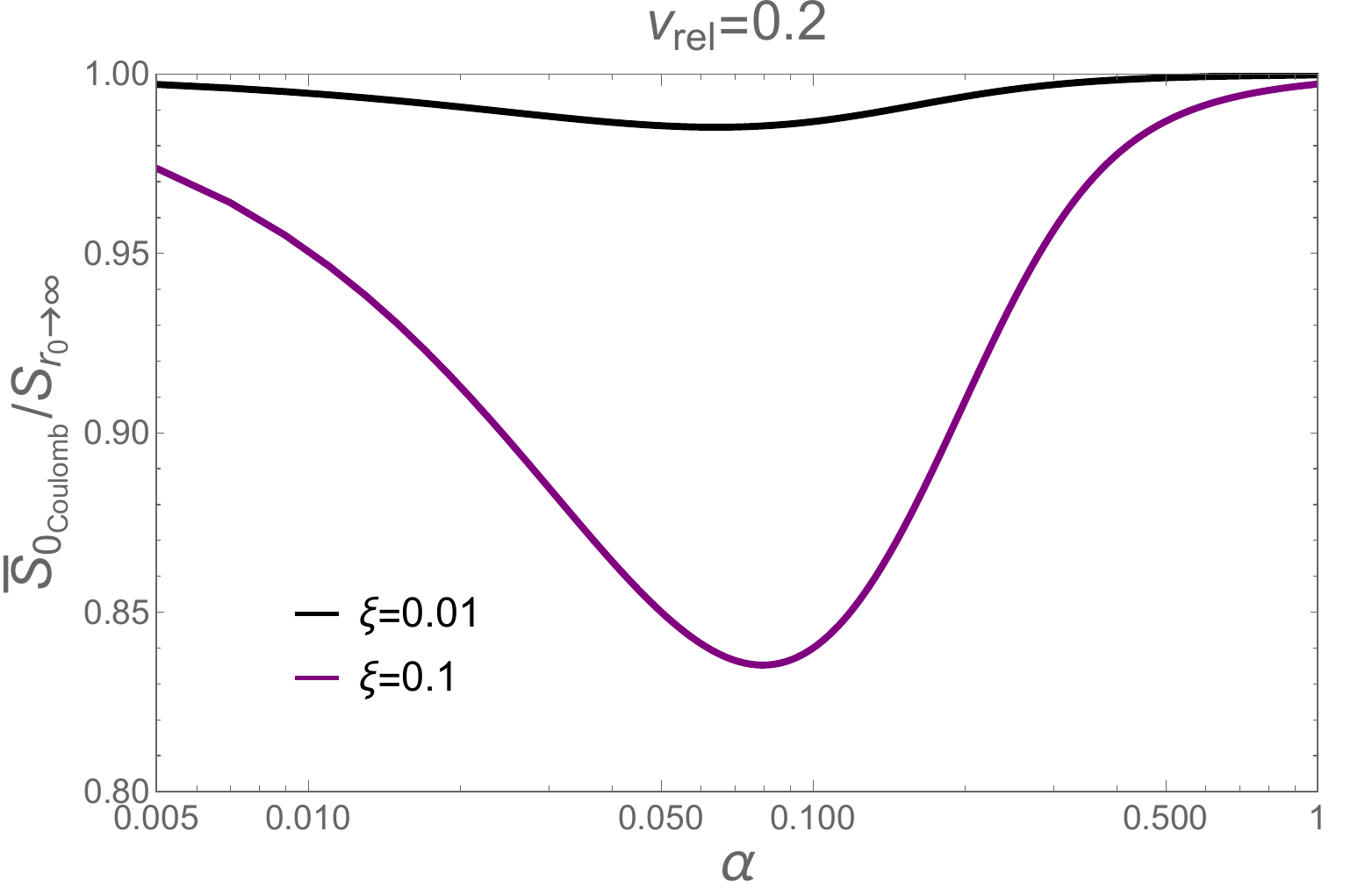} & 
\hspace{-0.2cm}
\includegraphics[height=5.8cm]{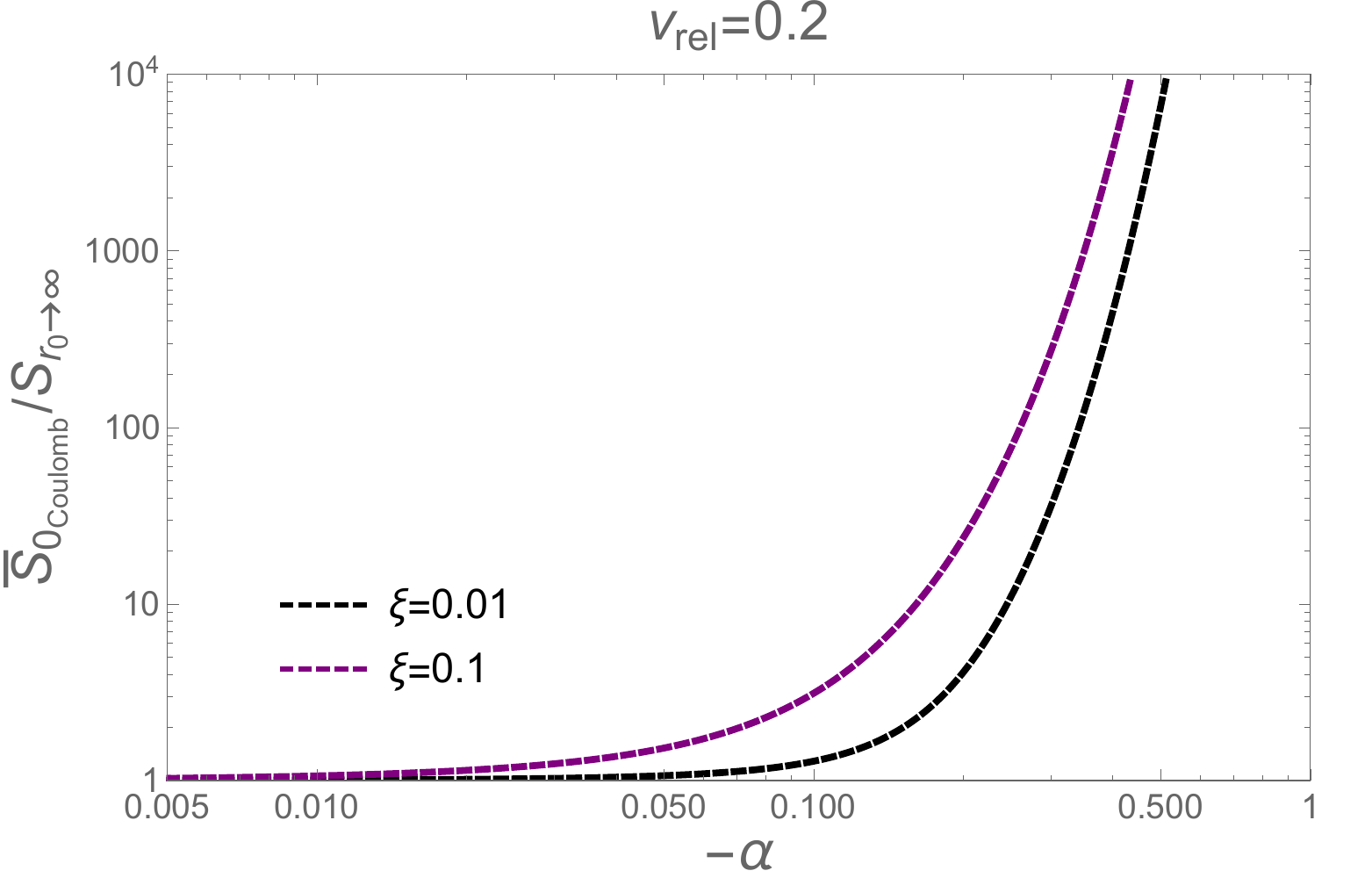} \\
\end{tabular}
\end{center}   
\caption{\label{fig:s-wave_Coulomb_bar}
\it 
Upper panels: the black and purple lines show $\bar{S}_{0_{Coulomb}}$ as functions of $v_{rel}$, for $\xi = 0.01$ and $0.1$, respectively. 
For comparison, $S_{r_0 \to \infty}$ is plotted using an orange line. 
Middle panels: the ratio of $\bar{S}_{0_{Coulomb}}$ to $S_{r_0 \to \infty}$, as a function of $\xi$.
The brown, pink and cyan lines are for $v_{rel} = 0.4$, $0.2$ and $0.1$, respectively. 
In both the upper and middle panels $|\alpha| = 0.1$ is used. 
Lower panels: the ratio of $\bar{S}_{0_{Coulomb}}$ to $S_{r_0 \to \infty}$, as a function of $\alpha$, for  
$v_{rel} = 0.2$. 
Again, the black and purple lines are for $\xi = 0.01$ and $0.1$, respectively.  
All left panels are for attractive Coulomb potentials where $\alpha > 0$, while all right panels are for repulsive ones where $\alpha < 0$.
}
\end{figure}

By introducing variables $\kappa \equiv \Gamma r_0/v_{rel}$ and $\xi \equiv 2 \Gamma/\mu$, the $\eta_0$ in $S_{0_{Coulomb}}$ can be written as $\eta_0 = 2 \kappa \alpha v_{rel} / \xi$. Then $\bar{S}_{0_{Coulomb}}$ becomes a function of $\xi, v_{rel}/\alpha$ and $\alpha v_{rel}$. 
Anticipating that in the next section we will average over $v_{rel}$ to get the thermally averaged Sommerfeld factor, we show in Figure~\ref{fig:s-wave_Coulomb_bar} $\bar{S}_{0_{Coulomb}}$ and $\bar{S}_{0_{Coulomb}}/S_{r_0 \to \infty}$ as functions of $v_{rel}$, $\alpha$ and $\xi$.

The left and right panels are for attractive and repulsive Coulomb potentials, respectively. 
In the upper panels, we choose $|\alpha| = 0.1$. 
The black and purple lines are for $\bar{S}_{0_{Coulomb}}$ with $\xi = 0.01$ and 0.1, respectively. 
The orange line is for $S_{r_0 \to \infty}$ with the same $\alpha$. 
The lines merge at large $v_{rel}$.
In the limit $v_{rel} \to \infty$, both $\bar{S}_{0_{Coulomb}}$ and $S_{r_0 \to \infty}$ go to $1$. 
In the other limit $v_{rel} \to 0$, $\bar{S}_{0_{Coulomb}}$ goes to $1$, while $S_{r_0 \to \infty}$ diverges 
for an attractive Coulomb potential and it exponentially vanishes for a repulsive one. 
Between these two limits, with the increase of $v_{rel}$, $\bar{S}_{0_{Coulomb}}$ first increases (decreases) and then decreases (increases) for an attractive (a repulsive) potential. On the contrary, $S_{r_0 \to \infty}$ monotonically changes. 
These behaviors can be understood intuitively. 
Without considering decays, when two particles in a pair approach each other with a slow relative velocity, they feel the Coulomb force for a long period of time before they meet and annihilate, so that the change of their wave function is large and thus the Sommerfeld enhancement or suppression is significant. 
However, the picture is different when particle decay is taken into account. 
For a small $v_{rel}$, the initial separation of the two particles in a pair has to be small, otherwise a decay is likely to occur before the two particles meet. 
If $v_{rel}$ is very small, the dominant contribution to $\bar{S}_{0_{Coulomb}}$ comes from pairs with small $|\eta_0|$, and we see from Eq.~(\ref{eq:r0_to_0_limit}) that $S_{0_{Coulomb}} \to 1$ as $|\eta_0| \to 0$.   
For a large $v_{rel}$, the two particles in a pair have a good chance to meet before a decay occurs even if their initial separation is not small. 
We can see that in general decay makes Sommerfeld factors less prominent compared to situations when the annihilating particles are stable. That is, when annihilating particle decays need to be considered, for a given set of $\alpha$ and $v_{rel}$, the enhancement factor is not that big for an attractive Coulomb potential, and the suppression factor is not that small for a repulsive one.  
The larger the decay rate is, the less prominent the Sommerfeld factor is. 
As we discussed in the Introduction, Sommerfeld effect becomes ineffective when $\sim (\mu v_{rel})^{-1}$ is larger than $\sim v_{rel}/\Gamma$. 
This explains qualitatively that in the upper panels the purple lines deviate from the orange lines at a larger $v_{rel}$, compared to the black lines.

Before we discuss other panels of Figure~\ref{fig:s-wave_Coulomb_bar}, let's pause to estimate the value of $\xi$ for a coannihilator pair.
For simplicity, consider that the coannihilator is a complex scalar $\tilde{C}$ with a mass $m$, and it decays into a Majorana DM $\chi$ with a mass $m_{\rm DM}$ and a Dirac fermion $f$ with a mass negligible compared to $\Delta m \equiv m - m_{\rm DM}$. 
From the Lagrangian $\mathcal{L} = c_d \tilde{C} \bar{\chi} f + h.c.$, where $c_d$ is a dimensionless coupling, one can get the decay rate $\Gamma_{\tilde{C}} = 2 \frac{|c_d|^2}{4\pi} \frac{(\Delta m)^2}{m} (1- \frac{\Delta m}{2 m})^2$. Therefore, $\xi = 8 \Gamma_{\tilde C}/m = 16 \frac{|c_d|^2}{4\pi} (\Delta m / m)^2 (1- \frac{\Delta m}{2 m})^2$. 
Up to the factor $\frac{|c_d|^2}{4\pi}$, $\xi$ is 0.0016, 0.014, 0.14 and 0.52 for $\Delta m / m = 0.01$, 0.03, 0.1 and 0.2, respectively. 
For a larger $\Delta m / m$, $\xi$ is larger, but usually coannihilation mechanism is ineffective. 
We consider $\frac{|c_d|^2}{4\pi} < 1$ for a perturbative coupling, and therefore we plot the range of $\xi$ from $0.001$ to $0.1$ in the middle panels, and we show cases of $\xi = 0.01$ and 0.1 in the upper and lower panels. 

Back to Figure~\ref{fig:s-wave_Coulomb_bar}, 
the middle panels show $\bar{S}_{0_{Coulomb}}/S_{r_0 \to \infty}$ as a function of $\xi$, for $|\alpha| = 0.1$ and three choices of $v_{rel}$, 0.4, 0.2 and 0.1. 
In sequence, these values of $v_{rel}$ are typical for annihilating pairs during freeze-out when the temperature $T$ decreases from $\sim m_{\rm DM} / 25$ to $\sim m_{\rm DM} / 100$ and to $\sim m_{\rm DM} / 400$. 
However, we should also keep in mind that $v_{rel}$ has a Maxwell-Boltzmann distribution for a given temperature. Therefore, for example, there are some pairs having $v_{rel} \sim 0.1$ or even smaller at $T \sim m_{\rm DM} / 25$. 
For each $v_{rel}$, $\bar{S}_{0_{Coulomb}}/S_{r_0 \to \infty}$ monotonically decreases (increases) from 1 with the increase of $\xi$, for an attractive (a repulsive) potential. 
For a given $\xi$, the change is larger for smaller $v_{rel}$. 
That is, the modification of the Sommerfeld factor due to decays of annihilating particles is more prominent for larger $\xi$ and/or smaller $v_{rel}$.

The lower panels show $\bar{S}_{0_{Coulomb}}/S_{r_0 \to \infty}$ as a function of $|\alpha|$ for $v_{rel} = 0.2$. 
The black and purple lines are for $\xi = 0.01$ and $\xi = 0.1$, respectively. 
In the limit $|\alpha| \to 0$, both $\bar{S}_{0_{Coulomb}}$ and $S_{r_0 \to \infty}$ go to 1. 
For attractive cases, $\bar{S}_{0_{Coulomb}}/S_{r_0 \to \infty}$ is close to 1 at large $\alpha$ for the choices of $v_{rel}$ and $\xi$, and it deviates from 1 the most at around $\alpha \sim 0.1$.
For repulsive cases, the deviation increases with the increase of $|\alpha|$. 
For both attractive and repulsive potentials, the deviations are bigger for larger $\xi$, since a larger decay rate makes the Sommerfeld enhancement or suppression less effective. 

\section{Modification of coannihilators' Sommerfeld effect during DM thermal freeze-out}
\label{sec:thermally_averaged_Sommerfeld_factor}

We are now in a position to consider coannihilators' decay effect on their Sommerfeld factor, and consequently on the DM thermal relic abundance. 
In this work we consider the simplest coannihilation scenario, in which there is only one species of DM particle $\chi$ and one species of coannihilator $\tilde{C}$. 
We assume that during freeze-out the interconversion rate between $\chi$ and $\tilde{C}$ is large enough in comparison to the Hubble expansion rate, so that for a given temperature the ratio of their number densities $\frac{n_\chi}{n_{\tilde{C}}}$ equals the equilibrium value $\frac{n^{eq}_\chi}{n^{eq}_{\tilde{C}}}$. 
This assumption is justified by our setup that we study cases in which the decay rate of $\tilde{C}$ into $\chi$ is much larger than the Hubble expansion rate. 
Therefore, the DM thermal relic abundance can be obtained by solving a single Boltzmann equation, 
\beq
\frac{dY}{dx}=-\frac{xs}{H(m_{\rm DM})}\left(1+\frac{T}{3g_{\ast s}}\frac{dg_{\ast s}}{dT}\right)\langle \sigma  v\rangle_{\rm eff} \left(Y^2- Y^2_{eq}\right) \,,
\label{eq:single_Boltzmann}
\eeq
in which $x$ is defined as the ratio of DM mass $m_{\rm DM}$ to temperature $T$, i.e., $x \equiv m_{\rm DM} / T$.  
The entropy density is
\beq
s = {2 \pi^2 \over 45} g_{\ast s} T^3 = {2 \pi^2 \over 45} g_{\ast s} m_{\rm DM}^3 / x^3 \,.
\eeq
$H(m_{\rm DM})$ is related with the Hubble expansion rate $H(T)$, as 
\beq
H(m_{\rm DM}) \equiv H(T) x^2 = \sqrt{4 \pi^3 g_\ast \over 45} {m_{\rm DM}^2 \over m_{\rm pl}} \,,
\eeq
where the Planck mass is $m_{\rm pl} \approx 1.22 \times 10^{19}$ GeV. 
$g_{\ast s}$ and $g_\ast$ are the total numbers of effectively massless degrees of freedom associated with the entropy density and the energy density of the thermal bath, respectively.
We assume that $\chi$ and $\tilde{C}$ have the same temperature as the Standard Model thermal bath. 
This can be achieved if interaction rates between dark sector particles and Standard Model sector particles are large enough in comparison to the Hubble expansion rate. 
Since we are not committed to a specific dark sector model, and the possibility that the two sectors have different temperatures is not important to the physics we want to focus on in this work, we take the simplest assumption. 
The yield $Y$ and its equilibrium value $Y_{eq}$ are defined as $Y \equiv \frac{n_\chi + n_{\tilde{C}}}{s}$ and $Y_{eq} \equiv \frac{n_\chi^{eq} + n_{\tilde{C}}^{eq}}{s}$, respectively. 
In order to maximize coannihilators' Sommerfeld effect on DM thermal relic abundance, in our calculation we neglect the (co)annihilation cross sections of $\chi \chi$ and $\chi \tilde{C}$. 
Therefore, the thermally averaged effective annihilation cross section (times relative velocity) is
\beq
\langle\sigma v \rangle_{\rm eff} = \langle\sigma_{\tilde{C} \tilde{C}} v_{rel} \rangle \frac{g_{\tilde{C}}^2 (1+\Delta m/m_{\rm DM})^3 e^{-2 x \Delta m / m_{\rm DM}}}{g_{\rm eff}^2} \, ,
\label{eq:sigma_eff}
\eeq
where $\Delta m$ is the mass difference between the coannihilator and the DM, i.e., $\Delta m \equiv m - m_{\rm DM}$. 
$g_{\rm eff}$ is given as
\beq
g_{\rm eff} \equiv g_\chi + g_{\tilde{C}} (1+\Delta m/m_{\rm DM})^{3/2} e^{- x \Delta m / m_{\rm DM}} \,,
\label{eq:g_eff}
\eeq
where $g_\chi$ and $g_{\tilde{C}}$ are the degrees of freedom of the DM particle and the coannihilator, respectively. 
$\sigma_{\tilde{C} \tilde{C}}$ is the usual spin-averaged (and also averaged over other intrinsic degrees of freedom, e.g., color, if applicable) cross section. 
If $\chi$ is not its own antiparticle, $g_\chi$ includes the contributions of both $\chi$ and $\bar{\chi}$. 
The same applies to $g_{\tilde{C}}$ as well, but the $\sigma_{\tilde{C} \tilde{C}}$ in Eq.~(\ref{eq:sigma_eff}) should then be replaced by $(\sigma_{\tilde{C} \tilde{C}} + \sigma_{\tilde{C} \bar{\tilde{C}}})/2$. A detailed explanation of the factor of $2$ can be found in the Appendix of~\cite{Srednicki:1988ce}. 
We assume that the number densities of particles and antiparticles are the same.
We will consider either $\langle\sigma_{\tilde{C} \bar{\tilde{C}}} v_{rel} \rangle$ or $\langle\sigma_{\tilde{C} \tilde{C}} v_{rel} \rangle$ dominates, and the Sommerfeld effect can be either an enhancement or a suppression. 
This means applying Eq.~(\ref{eq:sigma_eff}) to either the attractive or the repulsive case, no matter whether $\tilde{C}$ is its own antiparticle. 

By integrating Eq.~(\ref{eq:single_Boltzmann}) from a small $x$ when $Y = Y_{eq}$ to its value today which essentially corresponds to $x \to \infty$, we get today's yield, denoted by $Y_0$. 
The DM relic abundance $\Omega h^2$ is related with $Y_0$ as~\cite{Edsjo:1997bg}
\beq
\Omega h^2 = 2.755 \times 10^8 \frac{m_{\rm DM}}{\text {GeV}} Y_{0} \,.
\eeq

\subsection{Thermally averaged Sommerfeld factors for decaying coannihilators}

We consider $s$-wave annihilations. 
For a Coulomb potential,  
\beq
\langle\sigma_{\tilde{C} \tilde{C}} v_{rel} \rangle = 
a_{\tilde{C} \tilde{C}} \langle \bar{S}_{0_{Coulomb}}\rangle \,,
\label{eq:s_wave_annhilation_cross_section}
\eeq
where $a_{\tilde{C} \tilde{C}}$ is a constant, which is the $s$-wave value of $\sigma_{\tilde{C} \tilde{C}} v_{rel}$ without considering the Sommerfeld factor. 
$\langle \bar{S}_{0_{Coulomb}}\rangle$ is the thermally averaged Sommerfeld factor, given as  
\beq
\langle \bar{S}_{0_{Coulomb}}\rangle =
\int_0^{\infty} \bar{S}_{0_{Coulomb}} \Big(\frac{m}{4 \pi T}\Big)^{3/2} \, e^{\frac{-m v_{rel}^2}{4 T}} 4 \pi v_{rel}^2 d v_{rel} \,,
\label{eq:sigmavrel_thermal}
\eeq
where $\bar{S}_{0_{Coulomb}}$ is given in Eq.~(\ref{eq:sbar0Coulomb}), and it is a function of $\alpha$, $\xi$ and $v_{rel}$.  
Therefore, $\langle \bar{S}_{0_{Coulomb}}\rangle$ is a function of $\alpha$, $\xi$ and $m/T$. 

\begin{figure}
\begin{center}
\begin{tabular}{c c}
\hspace{-1.0cm}
\includegraphics[height=5.8cm]{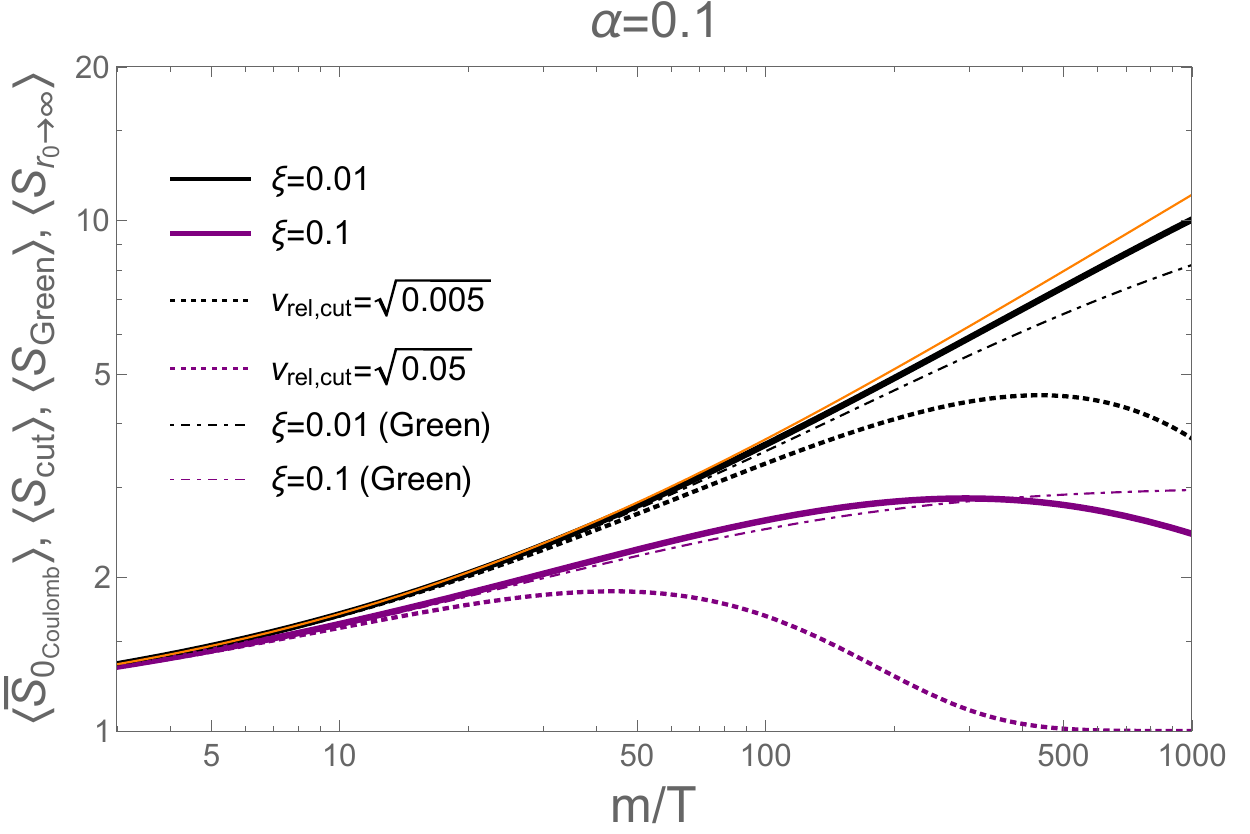} & 
\hspace{-0.2cm}
\includegraphics[height=5.8cm]{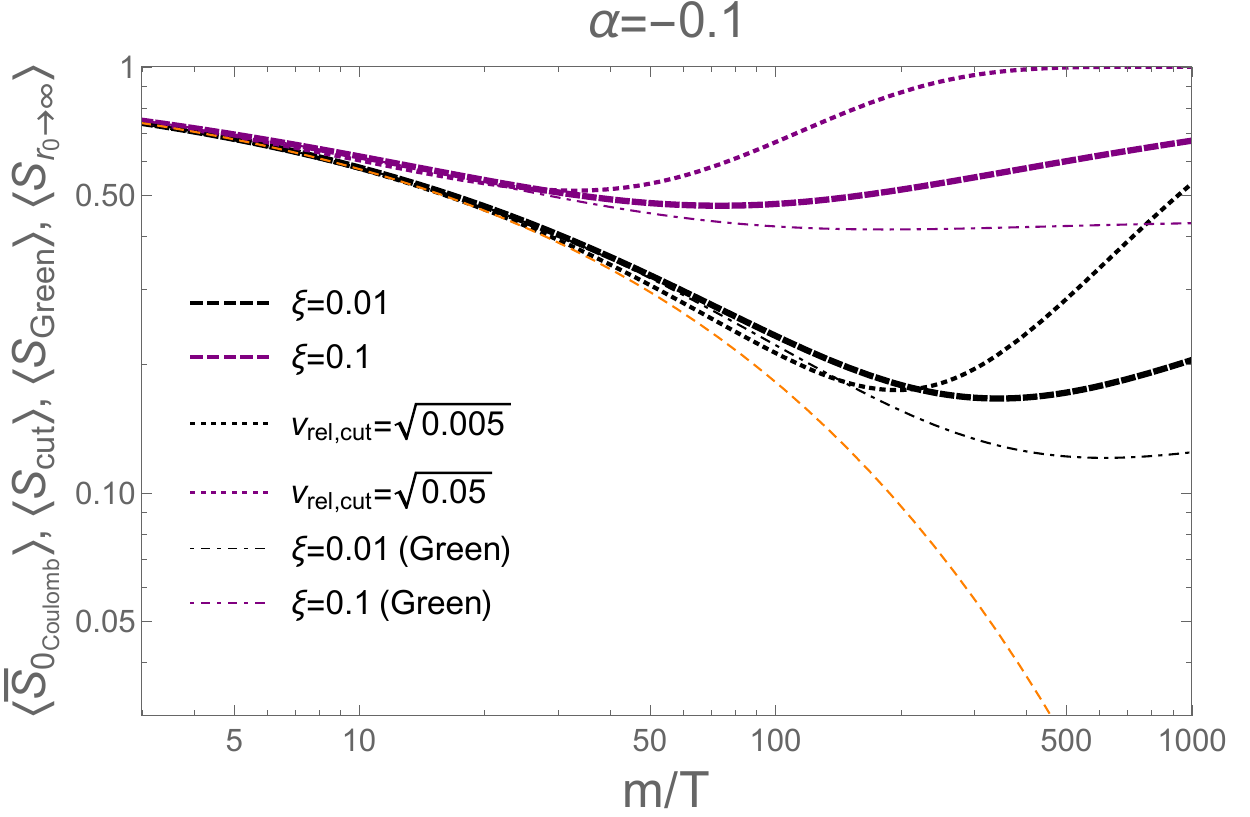} \\
\end{tabular}
\end{center}   
\caption{\label{fig:s-wave_Coulomb_bar_moverT}
\it
The thermally averaged $s$-wave Sommerfeld factors, $\langle \bar{S}_{0_{Coulomb}}\rangle$ (solid and dashed lines), $\langle S_{cut}\rangle$ (dotted lines) and $\langle S_{Green} \rangle$ (dot-dashed lines), for a Coulomb potential for a pair of unstable annihilating particles, as functions of the ratio of annihilating particle's mass to temperature. 
The black and purple lines are for $\xi = 0.01$ and $0.1$, respectively. 
For comparison, we use orange lines to plot $\langle S_{r_0 \to \infty}\rangle$, which is the thermally averaged $s$-wave Sommerfeld factor without considering decays of annihilating particles.  
The left panel is for an attractive potential where $\alpha = 0.1$, while the right panel is for a repulsive one where $\alpha = - 0.1$. 
}
\end{figure}

In Figure~\ref{fig:s-wave_Coulomb_bar_moverT} we plot $\langle \bar{S}_{0_{Coulomb}}\rangle$ as a function of $m/T$, using solid and dashed lines. 
The black and purple colors are for $\xi = 0.01$ and $0.1$, respectively. 
Since the lower-left panel of Figure~\ref{fig:s-wave_Coulomb_bar} shows that for an attractive Coulomb potential the modification of coannihilators' Sommerfeld factor reaches its maximum at around $\alpha \sim 0.1$, we choose $\alpha = 0.1$ in the left panel of Figure~\ref{fig:s-wave_Coulomb_bar_moverT}. 
For comparison, for the repulsive case we show $\alpha = - 0.1$ in the right panel, although we recall that the modification of coannihilators' Sommerfeld factor increases with the increase of $|\alpha|$, as can be seen in the lower-right panel of Figure~\ref{fig:s-wave_Coulomb_bar}. 
The solid and dashed orange lines show $\langle S_{r_0 \to \infty}\rangle$, which is the thermally averaged $s$-wave Sommerfeld factor without considering decays of coannihilators, that is, 
\beq
\langle S_{r_0 \to \infty}\rangle = \int_0^{\infty} S_{r_0 \to \infty} \Big(\frac{m}{4 \pi T}\Big)^{3/2} \, e^{\frac{-m v_{rel}^2}{4 T}} 4 \pi v_{rel}^2 d v_{rel} \,.
\label{eq:sigmavrel_thermal_infty}
\eeq
The orange line monotonically increases (decreases) with the increase of $m/T$ for the attractive (repulsive) case, since the typical $v_{rel}$ is smaller for larger $m/T$. 
The black and purple lines are closer to $1$ compared to the orange lines. 
One can see that when decays of coannihilators are considered, both the Sommerfeld enhancement and suppression are weaker. 
The modification of the thermally averaged Sommerfeld factor is more significant for larger $\xi$. 
In contrary to the monotonic behavior of the orange lines, the solid (dashed) black and purple lines first increase (decrease) with the increase of $m/T$, and then go to $1$ when $m/T$ is sufficiently large. 
This behavior was explained when we were discussing the upper panels of Figure~\ref{fig:s-wave_Coulomb_bar}. 
For the attractive case, the difference between the solid black (purple) line and orange line at $m/T = 25$ is about $1\%$ ($11\%$), while it becomes $4\%$ ($45\%$) at $m/T = 200$. 
For the repulsive case, the dashed black (purple) line is higher than the orange line by about $3\%$ ($24\%$) at $m/T = 25$, and by about a factor of 0.9 (4.5) at $m/T = 200$.

Before we compute the DM thermal relic abundance, let's pause to compare the thermally averaged Sommerfeld factor obtained in this work with the ones calculated by two other methods. 

The first one is the velocity-cut method used in our previous work~\cite{Cui:2020ppc}. 
In that method, the thermally averaged $s$-wave Sommerfeld factor taking into account coannihilators' decay is given as, by using the notation in the current work,
\beq
\langle S_{cut}\rangle \equiv \int_0^{\infty} \Big(\frac{m}{4 \pi T}\Big)^{3/2} \, e^{\frac{-m v_{rel}^2}{4 T}} 4 \pi v_{rel}^2 [(S_{r_0 \to \infty} - 1) H (v_{rel} - v_{cut}) + 1]  d v_{rel} \,,
\label{eq:sigmavrel_thermal_cut}
\eeq
where $H (v_{rel} - v_{cut})$ is the Heaviside step function, and the cut-off velocity  $v_{cut}$ is equal to $\sqrt{\xi/2}$. 
We plot $\langle S_{cut}\rangle$ using dotted lines in Figure~\ref{fig:s-wave_Coulomb_bar_moverT}. 
By comparing the dotted lines with the solid or dashed lines with the same color, we can see that the general behaviors of the curves are same. 
Also, curves with the same color are close at small $m/T$, while the velocity-cut method gives a larger effect at large $m/T$. 
It indicates that the results in the current work are more conservative. 

The second one is the Green's function approach based on non-relativistic
quantum field theory. 
The $s$-wave Sommerfeld factor can be written as (see Eq.~(4.8) of~\cite{deLima:2022joz}), adapted to our notation, 
\beq
S_{Green} = \frac{{\rm Im} G(\frac{1}{2} \mu v_{rel}^2 + i \frac{\Gamma}{2}, \vec{r}, 0)|_{\vec{r} \to 0}}{{\rm Im} G_0(\frac{1}{2} \mu v_{rel}^2 + i \frac{\Gamma}{2}, \vec{r},0)|_{\vec{r} \to 0}} \,.
\eeq 
The free Green's function is given by 
\beq
G_0(\frac{1}{2} \mu v_{rel}^2 + i \frac{\Gamma}{2}, \vec{r},0) = \frac{2 \mu}{4 \pi r} e^{i \sqrt{2 \mu (\frac{1}{2} \mu v_{rel}^2 + i \frac{\Gamma}{2})} \, r} = \frac{2 \mu}{4 \pi r} e^{i \mu r \sqrt{v_{rel}^2 + i \frac{\xi}{2}}} \,. 
\eeq
The Coulomb Green's function is~\cite{10.1063/1.1704153}
\beq
G(\frac{1}{2} \mu v_{rel}^2 + i \frac{\Gamma}{2}, \vec{r},0) =  \frac{2 \mu}{4 \pi r} \Gamma (1- i \nu) W_{i \nu , \frac{1}{2}} (- 2 i \mu r \sqrt{v_{rel}^2 + i \frac{\xi}{2}} \, )
\,, 
\eeq
where the $\nu$ appearing in the argument of the Gamma function $\Gamma (1- i \nu)$ is 
$\nu \equiv \frac{\alpha \mu}{\sqrt{2 \mu (\frac{1}{2} \mu v_{rel}^2 + i \frac{\Gamma}{2})}} = \frac{\alpha}{\sqrt{v_{rel}^2 + i \frac{\xi}{2}}}$, and $W_{i \nu , \frac{1}{2}} (- 2 i \mu r \sqrt{v_{rel}^2 + i \frac{\xi}{2}}\,)$ is the Whittaker function. 
It can be checked that when $\xi = 0$, $S_{Green} = S_{r_0 \to \infty}$. 
In Figure~\ref{fig:s-wave_Coulomb_bar_moverT}, we plot using dot-dashed lines the thermally averaged $S_{Green}$, 
\beq
\langle S_{Green} \rangle = \int_0^{\infty} S_{Green} \Big(\frac{m}{4 \pi T}\Big)^{3/2} \, e^{\frac{-m v_{rel}^2}{4 T}} 4 \pi v_{rel}^2 d v_{rel} \,.
\eeq
We can see that the general behaviors of the dot-dashed lines are also similar to the corresponding solid or dashed lines with the same color.  
In particular, for the range of not very large $m/T$, where it is most relevant for the calculation of the DM thermal relic abundance in the coannihilation scenarios we are considering, the dot-dashed lines and the corresponding solid or dashed lines are very close. 

These comparisons strengthen the viability of the current method, and can serve to verify our main finding that the coannihilators' decay makes the Sommerfeld enhancement or suppression less effective. 

\subsection{Effect on the DM thermal relic abundance}

The relative change of the DM thermal relic abundance due to the modification of the Sommerfeld factor is denoted by $\Delta \Omega/\Omega_{r_0 \to \infty}$, which is defined as
\beq
\Delta \Omega/\Omega_{r_0 \to \infty} \equiv \Omega h^2/(\Omega h^2)_{r_0 \to \infty} -1 \,, 
\label{eq:delta_Omega}
\eeq
where $(\Omega h^2)_{r_0 \to \infty}$ is $\Omega h^2$ but with $ \langle \bar{S}_{0_{Coulomb}}\rangle$ substituted by $\langle S_{r_0 \to \infty}\rangle$ in Eq.~(\ref{eq:s_wave_annhilation_cross_section}). 

\begin{figure}
\begin{center}
\includegraphics[height=7.cm]{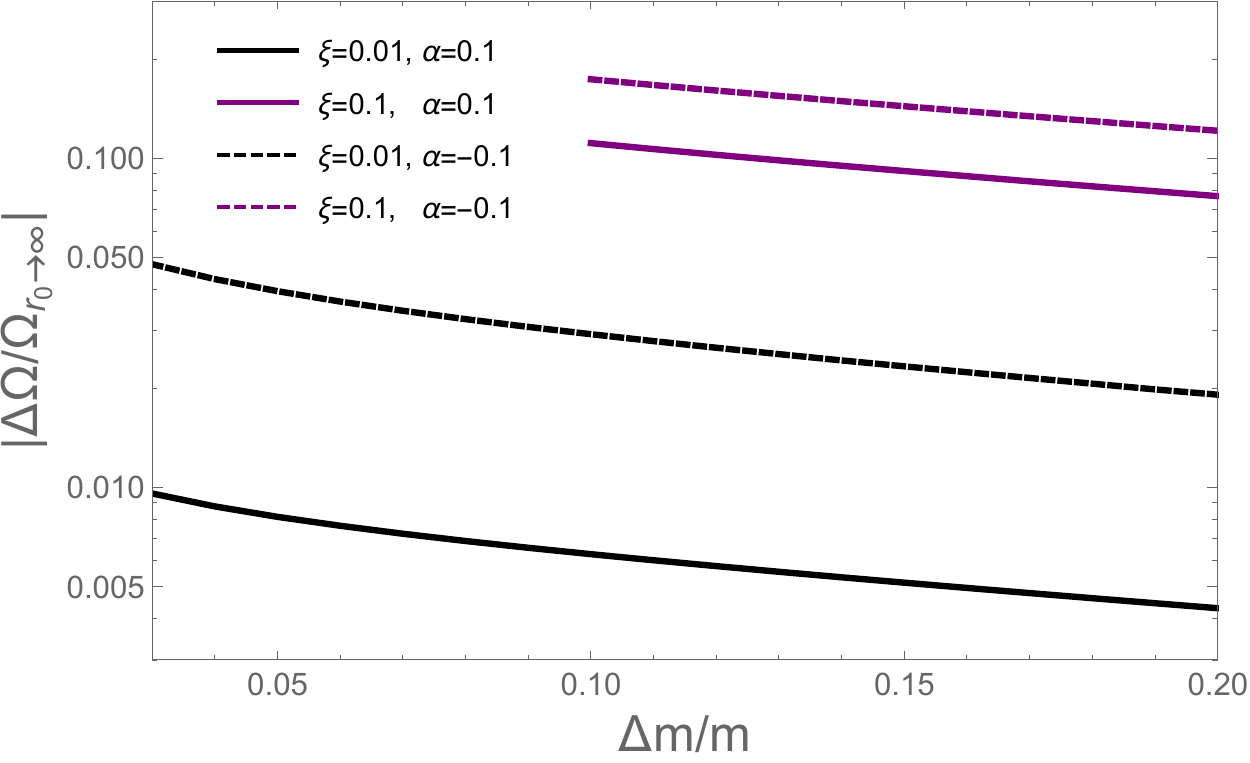}
\end{center}   
\caption{\label{fig:Delta_ohsq}
\it
The relative change of the DM thermal relic abundance due to the modification of the Sommerfeld factor induced by coannihilators' decay, $\Delta \Omega/\Omega_{r_0 \to \infty}$, as a function of $\Delta m/m$.
$\Delta \Omega/\Omega_{r_0 \to \infty}$ is positive for an attractive Coulomb potential, while it is negative for a repulsive one. 
The black and purple lines are for $\xi = 0.01$ and $0.1$, respectively. The solid lines are for an attractive potential with $\alpha = 0.1$, while the dashed lines are for a repulsive one with $\alpha = - 0.1$. 
}
\end{figure}

Due to the exponential factor in Eq.~(\ref{eq:sigma_eff}), coannihilation mechanism becomes ineffective for large $x$ if $\Delta m/m$ is not very small. 
On the other hand, the modification of the Sommerfeld factor is negligible if $\Delta m/m$ is too small, because a small $\Delta m/m$ cannot give a sizable $\xi$. 
Therefore, to study $\Delta \Omega/\Omega_{r_0 \to \infty}$ in the simple coannihilation scenario, in Figure~\ref{fig:Delta_ohsq} we consider $\Delta m/m$ between 0.03 and 0.2 for $\xi = 0.01$, and $\Delta m/m$ between 0.1 and 0.2 for $\xi = 0.1$.
These ranges of $\Delta m/m$ and the corresponding $\xi$ also ensure that the coupling between $\tilde{C}$ and $\chi$ is perturbative, for the simple model we discussed in section~\ref{sec:finite-range_averaged_Sommerfeld_factor}.  
Considering that $x = (1-\Delta m/m) (m/T)$, and since $\Omega h^2$ is approximately inversely proportional to $\langle\sigma v \rangle_{\rm eff}$, we can estimate from Figure~\ref{fig:s-wave_Coulomb_bar_moverT} that $|\Delta \Omega/\Omega_{r_0 \to \infty}|$ should be of order $\mathcal{O}(1\%)$ for $\xi = 0.01$ and $\mathcal{O}(10\%)$ for $\xi = 0.1$. 

In Figure~\ref{fig:Delta_ohsq}, for $\alpha = \pm  0.1$ and $\xi = 0.01$ or $0.1$, we plot $|\Delta \Omega/\Omega_{r_0 \to \infty}|$ as a function of $\Delta m/m$, for a choice of parameters  $g_\chi = g_{\tilde{C}} = 2$, $m_{\rm DM} = 2 \times 10^3 \,\text{GeV}$ and $a_{\tilde{C} \tilde{C}} = 10^{-8} \, \text{GeV}^{-2}$. 
$\Delta \Omega/\Omega_{r_0 \to \infty}$ is positive for $\alpha = 0.1$, while it is negative for $\alpha = - 0.1$. 
$|\Delta \Omega/\Omega_{r_0 \to \infty}|$ decreases with the increase of $\Delta m/m$, since coannihilation mechanism is less effective for larger $\Delta m/m$. 
On each line since $\xi$ is fixed, the coupling between $\tilde{C}$ and $\chi$ is smaller for larger $\Delta m/m$. 
We can see that for each line $|\Delta \Omega/\Omega_{r_0 \to \infty}|$ is close to the difference between $\langle \bar{S}_{0_{Coulomb}}\rangle$ and $\langle S_{r_0 \to \infty}\rangle$ at $m/T \sim 25$ in Figure~\ref{fig:s-wave_Coulomb_bar_moverT}, indeed as we have estimated. 

\begin{figure}
\begin{center}
\begin{tabular}{c c}
\hspace{-1.0cm}
\includegraphics[height=9.cm]{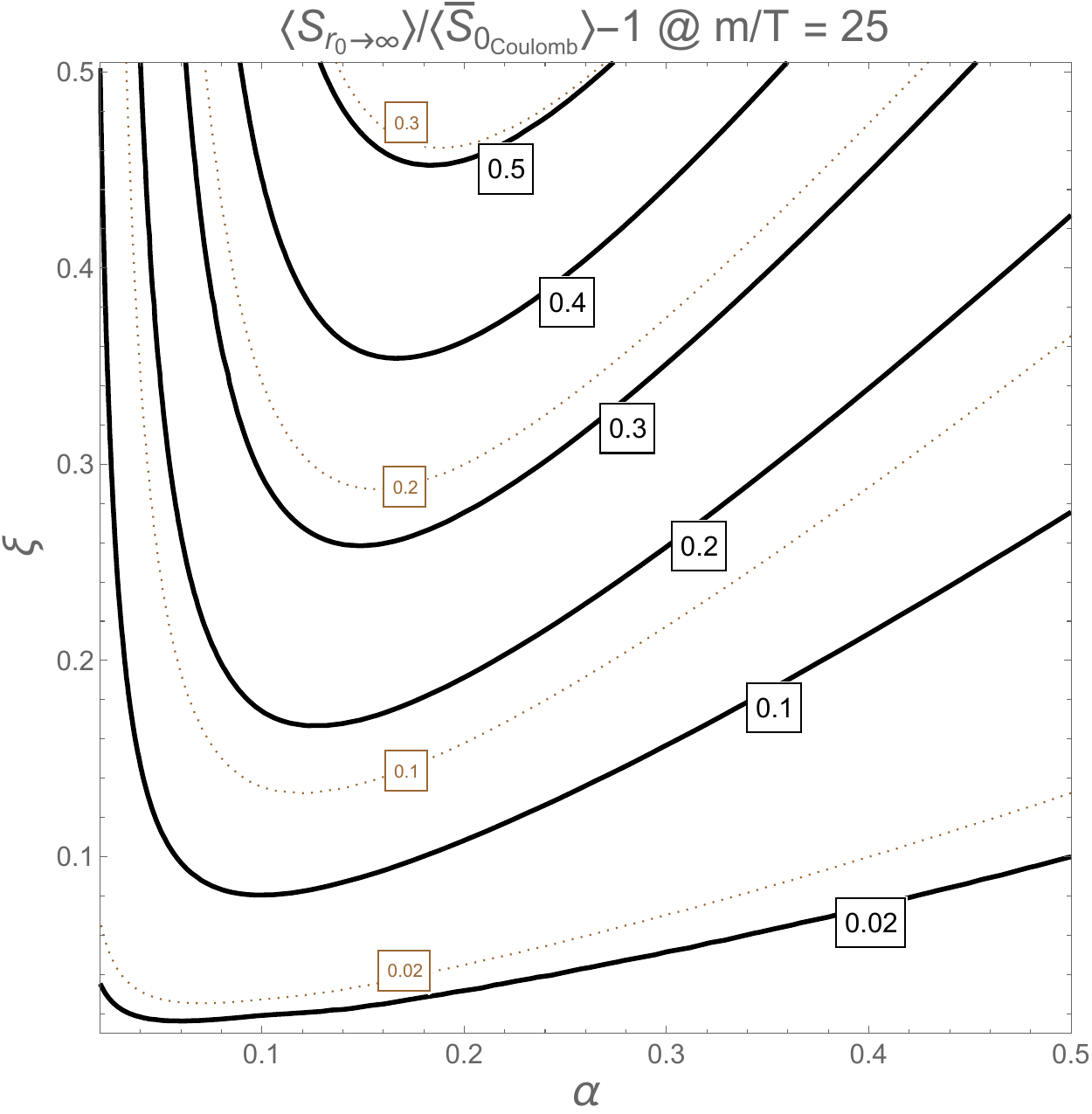} & 
\hspace{-0.2cm}
\includegraphics[height=9.cm]{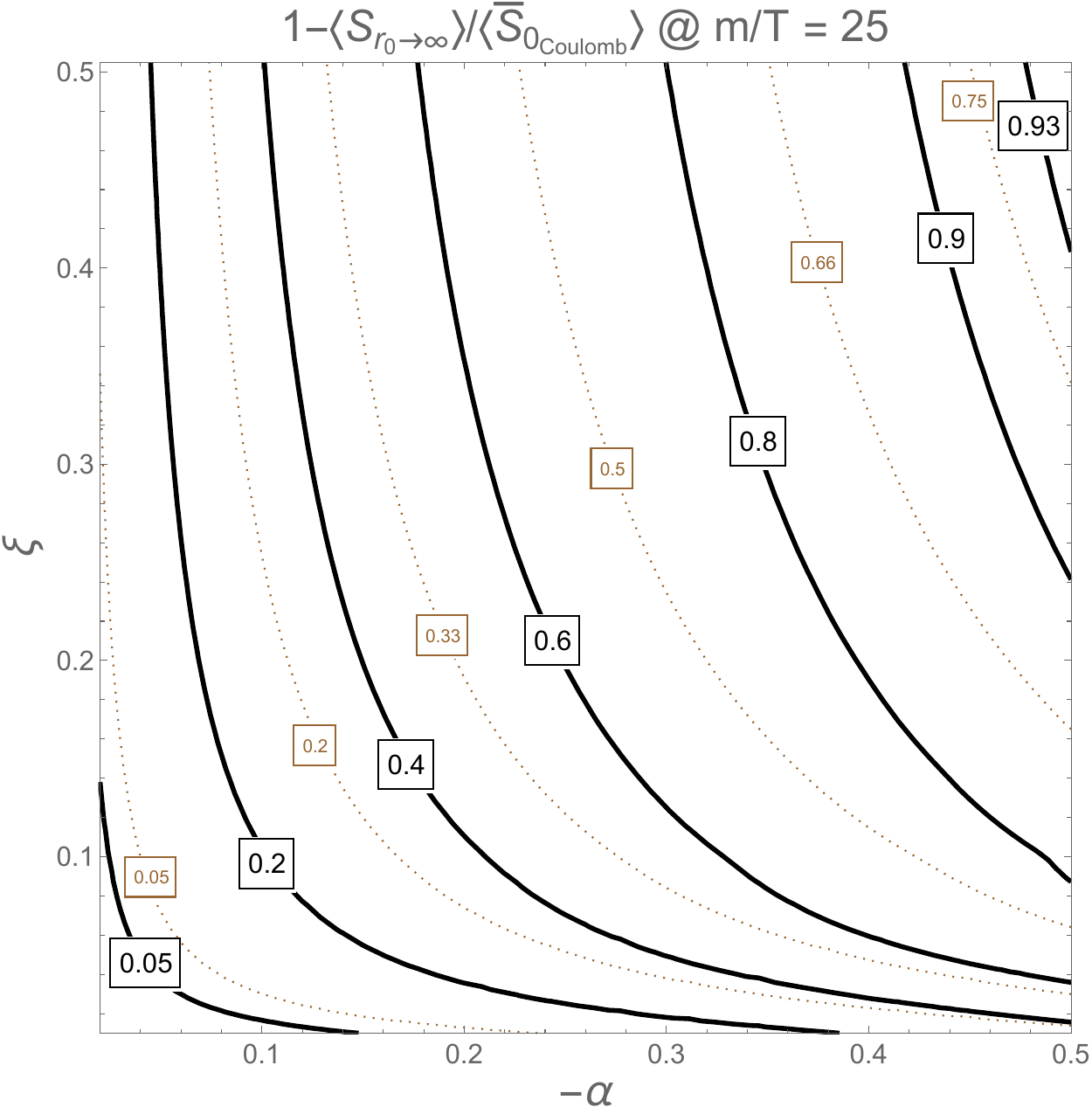} \\
\end{tabular}
\end{center}   
\caption{\label{fig:Coulomb_scan}
\it
The black solid lines are $|\langle S_{r_0 \to \infty}\rangle / \langle \bar{S}_{0_{Coulomb}}\rangle  - 1|$ contours computed at $m/T = 25$ for Coulomb potentials. 
The brown dotted lines are $|\Delta \Omega/\Omega_{r_0 \to \infty}|$ contours computed for $\Delta m / m = 0.2$ using $g_\chi = g_{\tilde{C}} = 2$, $m_{\rm DM} = 2 \times 10^3 \,\text{GeV}$ and $a_{\tilde{C} \tilde{C}} = 10^{-8} \, \text{GeV}^{-2}$. 
The left (right) panel is for attractive (repulsive) cases, where $\langle S_{r_0 \to \infty}\rangle / \langle \bar{S}_{0_{Coulomb}}\rangle  - 1$ and $\Delta \Omega/\Omega_{r_0 \to \infty}$ are positive (negative). 
}
\end{figure}

To obtain an estimate of the potential magnitude of the effect on the DM thermal relic abundance, in Figure~\ref{fig:Coulomb_scan} we compute $|\langle S_{r_0 \to \infty}\rangle / \langle \bar{S}_{0_{Coulomb}}\rangle  - 1|$ at $m/T = 25$ on the ($\alpha, \xi$) plane. 

For both attractive and repulsive cases, for a given $\alpha$ the values of contours are larger for larger $\xi$, meaning that a larger decay rate makes the Sommerfeld enhancement or suppression less effective. 
For a given $\xi$, for the attractive case the values of contours first become larger and then become smaller with the increase of $\alpha$, while for the repulsive case the values monotonically increase with the increase of $|\alpha|$. 
These behaviors can be also found in the lower panels of Figure~\ref{fig:s-wave_Coulomb_bar}. 
It is due to the relative size of two length scales, namely, the initial separation of a pair of annihilating coannihilators and the Bohr radius. 
The modification of the Sommerfeld factor is significant when the former is comparable or smaller than the latter. 
The former decreases with the increase of $\xi$, while the latter is inversely proportional to $\alpha$. 
On the other hand, for large $\alpha$, while for the attractive case $S_{r_0 \to \infty}$ increases proportionally with the increase of $\alpha$, for the repulsive case it decreases exponentially with the increase of $|\alpha|$. 

For an attractive Coulomb potential, the difference between $\langle \bar{S}_{0_{Coulomb}}\rangle$ and $\langle S_{r_0 \to \infty}\rangle$ can be as much as $\sim 50\%$ for $\alpha \sim 0.2$ and $\xi \sim 0.46$. 
We recall that for the simple model we discussed in section~\ref{sec:finite-range_averaged_Sommerfeld_factor}, in order to maintain a perturbative coupling between $\tilde{C}$ and $\chi$, $0.52$ is the largest value that $\xi$ can take for $\Delta m / m = 0.2$. 
In Figure~\ref{fig:Coulomb_scan}, we also show using brown dotted lines $|\Delta \Omega/\Omega_{r_0 \to \infty}|$ contours computed for $\Delta m / m = 0.2$ and the same choice of parameters as in Figure~\ref{fig:Delta_ohsq}, namely, $g_\chi = g_{\tilde{C}} = 2$, $m_{\rm DM} = 2 \times 10^3 \,\text{GeV}$ and $a_{\tilde{C} \tilde{C}} = 10^{-8} \, \text{GeV}^{-2}$. 
The solid black and dotted brown contours have the same features and they differ by less than a factor of $2$. 
It double confirms the viability to use $|\langle S_{r_0 \to \infty}\rangle / \langle \bar{S}_{0_{Coulomb}}\rangle  - 1|$ at $m/T = 25$ as a reasonable estimate of the effect of the modified Sommerfeld factor on the DM thermal relic abundance. 

We conclude that, when there is an attractive Coulomb-like force between a pair of annihilating coannihilators, the modification of the $s$-wave Sommerfeld factor induced by coannihilators' decay can potentially increase the calculated DM thermal relic abundance by as much as several tens of percent; when the force is repulsive, the calculated DM thermal relic abundance can be reduced by a factor of a few.

We note that the modification of the coannihilators' Sommerfeld factor is determined by $\alpha$ and $\xi$. 
Other quantities, such as the DM mass and the DM-DM into Standard Model annihilation cross sections, determine the DM phenomenology, but have little influence on the coannihilators' Sommerfeld factor. 
Nevertheless, since $\xi$ is closely related to $\Delta m$, which is a critical parameter in collider search of DM in coannihilation scenarios (for instance, jets plus missing transverse energy searches), in complete BSM models the modification of the Sommerfeld factor may lead to a shift of the parameter regions which can both give correct DM thermal relic abundance and be testable by collider experiments.

\section{Summary}
\label{sec:summary}

We have calculated Sommerfeld factors for a pair of unstable annihilating particles.  
Due to decays, the two particles have to approach each other from a finite initial separation, from where they start to feel the long-range potential generated by themselves.
Consequently, conventional calculations of Sommerfeld factors which essentially assume an infinite initial separation may need to be modified. 

To illustrate the physics, we focus our discussions on the $s$-wave Sommerfeld factor for a truncated Coulomb potential. We use the truncation distance to take into account the information that the initial separation of the two annihilating particles is finite. 
This distance is then averaged over accounting for the probabilistic nature of decays. 
The resultant decay-rate-averaged Sommerfeld factors (RASFs) show that Sommerfeld effects are less prominent compared to situations when the annihilating particles are stable. 
The modifications are more significant for larger decay rates and/or smaller relative velocities. 
This confirms our intuitive idea. 
For an annihilation to happen, the typical initial separation of two incoming particles is given by the ratio of their relative velocity to the sum of their decay rates. 
Large decay rates and/or a small relative velocity lead to a small initial separation, so that the accumulation of the changes of the two-body wave function from a plane wave is small, and consequently the Sommerfeld effect is less effective. 

Using the RASFs, we study thermally averaged $s$-wave Sommerfeld factors for a pair of unstable annihilating particles. 
Applying the result to a simple coannihilation scenario, we find that the modification of annihilating coannihilators' Sommerfeld factors caused by coannihilator decays may lead to a change of the DM thermal relic abundance well beyond the percent level.  

Before we close, we note that there are other approaches to compute the Sommerfeld factor for unstable particles~\cite{hep-ph/9501214,deLima:2022joz,Matsumoto:2022ojl}, in addition to the method based on the scattering wave function in the non-relativistic quantum mechanics framework, which we used in this work. It would be interesting to develop those approaches in the context of coannihilation scenarios, where the unstable coannihilator is customarily taken to be on-shell. This is different in the collider situations, where the unstable final state particles are usually taken to be off-shell in the calculations of their Sommerfeld factors. 
Finally, in the parameter region where the coannihilator and DM are very degenerate in mass, the coannihilator (into DM) scattering rate, rather than the coannihilator (into DM) decay rate, dominates the coannihilator-DM interconversion rate. The investigations of this scenario will be left for future work. 

\section*{Acknowledgments}
The author thanks Xiaoyi Cui, Yuangang Deng, Michihisa Takeuchi and Zhenhua Yu for helpful discussions. This work is supported by the Sun Yat-sen University Science Foundation. 

\appendix
\section{Sommerfeld factors for a generic finite-range central-force potential}
\label{sec:appendix A}
Techniques for calculating the Sommerfeld factor are available in the literature (see e.g. \cite{Arkani-Hamed:2008hhe, Iengo:2009ni, Cassel:2009wt, Slatyer:2009vg}). 
In this appendix, we present in a pedagogical approach the procedure in obtaining the $l$-wave Sommerfeld factor for a generic finite-range central-force potential $V(r)$, meaning that $V(r)$ vanishes for $r > r_0$.  
We require that the potential satisfies $r^2 V(r) \to 0$ for $r \to 0$. 
This includes the widely used Coulomb, Yukawa and Hulth\'en potentials. 
After deriving general formulae, we give explicit expressions for a finite-range Coulomb potential, which we use in the main text.  

Suppose that the long-range interaction between two massive annihilating particles can be described by a finite-range central-force potential $V(r)$, and that the annihilation happens at $r = 0$, 
the Sommerfeld factor can be determined by solving for the scattering wave function of the Schr\"odinger equation for the relative motion, 
\beq
\Big[- \frac{1}{2 \mu} \nabla_{\vec{r}}^2 + V(r) \Big] \psi (\vec{r}) = E \psi (\vec{r}) \,,
\label{eq:one_particle_schrodinger}
\eeq
where we have set $\hbar \equiv 1$. $\mu$ is the reduced mass of the two-particle system. 
$E$ is related with the relative momentum $\vec{k}$ and the relative velocity $v_{rel}$ of the two incoming particles at large separation when $V(r) = 0$, satisfying $k \equiv |\vec{k}| = \sqrt{2 \mu E} = \mu v_{rel}$. 
Because of the axial symmetry about the $z$-axis, which is the direction of the incoming particles at large distance, the solution $\psi (\vec{r})$ takes the form 
\beq
\psi(\vec{r}) \equiv  \sum_{l = 0}^\infty \psi_l(\vec{r}) = \sum_{l = 0}^\infty A_l P_l (\cos\theta) R_{kl} (r)  \,,
\label{eq:psi_separation_of_variables}
\eeq 
where $A_l$ are constants and $P_l (\cos\theta)$ are the Legendre polynomials. $\theta$ is the angle between $\vec{r}$ and the $z$-axis. $R_{kl} (r)$ are the radial  functions associated with the orbital angular momentum quantum number $l$, and these functions are real. 

Working with spherical coordinates, using 
\beq
\nabla_{\vec{r}}^2 = \frac{2}{r} \frac{\partial}{\partial r} + \frac{\partial^2}{\partial r^2} + \frac{1}{r^2} \Big[\frac{1}{\sin \theta} \frac{\partial}{\partial \theta} \Big(\sin\theta \frac{\partial}{\partial \theta}\Big) + \frac{1}{\sin^2 \theta} \frac{\partial^2}{\partial \phi^2} \Big] \equiv \frac{2}{r} \frac{\partial}{\partial r} + \frac{\partial^2}{\partial r^2} - \frac{\hat{l}^2}{r^2}  \,,
\eeq
\beq
\hat{l}^2 P_l (\cos\theta) = l (l+1) P_l (\cos\theta) \,,
\eeq
and the orthogonal relation 
\beq
\int_{-1}^{1} P_l (\cos\theta) P_{l^\prime} (\cos\theta) d (\cos \theta) = \frac{2}{2 l + 1} \delta_{l l^\prime}  \,, 
\label{eq:Legendre_orthogonal}
\eeq
for each $l$ Eq.~(\ref{eq:one_particle_schrodinger}) gives
\beq
\frac{d^2 R_{kl} (r)}{dr^2} + \frac{2}{r} \frac{d R_{kl} (r)}{dr} + \Big[k^2 - \frac{l(l+1)}{r^2} - 2 \mu V(r) \Big] R_{kl} (r) = 0 \,.
\eeq

One can solve for $R_{kl} (r)$ for both the ranges of $r < r_0$ and $r > r_0$. 
In either range the solution has two constants. 
The total four constants are determined by the following four conditions.
The requirement that $R_{kl} (r)$ is finite as $r \to 0$ gives one condition, 
\beq
R_{kl} (r) \propto r^l \;\; \text{as} \;\; r \to 0 \,.
\label{eq:solution_condition_1}
\eeq
In the range $r > r_0$, $V(r) = 0$, and $R_{kl} (r)$ takes the form 
\beq
R_{kl} (r) = c_{l_{\rm outer1}} j_l (kr) + c_{l_{\rm outer2}}  y_l (kr) \,, \;\; (\text{for} \;\; r > r_0) \,,
\label{eq:outer_solution}
\eeq
where $j_l (kr)$ and $y_l (kr)$ are spherical Bessel functions of the first and second kind, respectively. $c_{l_{\rm outer1}}$ and $c_{l_{\rm outer2}}$ are real constants.
The asymptotic form of $R_{kl} (r)$ at $r \to \infty$ is 
\beq
R_{kl} (r) \overset{r \to \infty}{\longrightarrow} \frac{1}{kr} \Big[c_{l_{\rm outer1}} \sin (kr - \frac{l \pi}{2}) - c_{l_{\rm outer2}} \cos (kr - \frac{l \pi}{2}) \Big] \,.
\eeq
One can choose to normalize $R_{kl} (r)$, such that
\beq
R_{kl} (r) \overset{r \to \infty}{\longrightarrow} \frac{2}{r} \sin (kr - \frac{l \pi}{2} + \delta_l) \,,
\eeq
where $\delta_l$ is the phase shift, which is real.   
The normalization gives the relations $c_{l_{\rm outer1}} = 2k \cos\delta_l$ and $c_{l_{\rm outer2}} = - 2k \sin\delta_l$, so that Eq.~(\ref{eq:outer_solution}) becomes
\beq
R_{kl} (r) = 2k \cos \delta_l j_l (kr) - 2k \sin \delta_l  y_l (kr) \,, \;\; (\text{for} \;\; r > r_0)  \,.
\label{eq:outer_solution_normalized}
\eeq
This is the second condition.
The third and fourth conditions are that $R_{kl} (r)$ and $\frac{d R_{kl} (r)}{dr}$ are continuous at $r = r_0$. 

The $l$-wave Sommerfeld factor is 
\beq
S_l = \lim_{r \to 0} \Big\lvert \frac{\psi_l (r)}{\psi_{l,free} (r)} \Big\rvert^2 
= \Big\lvert \frac{A_l}{A_{l,free}} \Big\rvert^2 \lim_{r \to 0} \Big\lvert \frac{R_{kl} (r)}{R_{kl, free} (r)} \Big\rvert^2  \,,
\label{eq:l-wave_Sommerfeld}
\eeq
where $\psi_{l,free} (r)$ is the $l$-wave function without the potential term for all $r$ in Eq.~(\ref{eq:one_particle_schrodinger}), and it can also take the form of Eq.~(\ref{eq:psi_separation_of_variables}),
\beq
\psi_{free} (\vec{r}) \equiv  \sum_{l = 0}^\infty \psi_{l,free} (\vec{r}) = \sum_{l = 0}^\infty A_{l,free} P_l (\cos\theta) R_{kl,free} (r)  \,.
\label{eq:psi_free}
\eeq 

We can get $A_l$ by the standard method in scattering theory (see e.g. \cite{Landau:1991wop}). 
Making use of   
\beq
e^{ikz} \overset{r \to \infty}{\longrightarrow} \frac{1}{2ikr} \sum_{l=0}^\infty (2l+1) P_l (\cos\theta) \big[e^{ikr} + (-1)^{l+1} e^{-ikr} \big] \,,
\eeq
\beq
\sin (kr - \frac{l \pi}{2} + \delta_l) = \frac{1}{2i} \big[e^{ikr}e^{i(- \frac{l \pi}{2} + \delta_l)} - e^{-ikr}e^{-i(-\frac{l \pi}{2} + \delta_l)} \big] \,,
\eeq
and Eq.~(\ref{eq:Legendre_orthogonal}), 
and comparing the coefficients of $e^{ikr}$ and $e^{-ikr}$ of the two asymptotic forms of $\psi (r)$,  
\beq
\psi (r) \overset{r \to \infty}{\longrightarrow} e^{ikz} + f(\theta) \frac{e^{ikr}}{r} 
\eeq 
and 
\beq
\psi (r) \overset{r \to \infty}{\longrightarrow} \sum_{l=0}^\infty A_l P_l (\cos\theta) \frac{2}{r} \sin (kr - \frac{l \pi}{2} + \delta_l) \,,
\eeq
we get 
\beq
A_l = \frac{1}{2 k} (2l+1) i^l e^{i \delta_l} \,.
\eeq
The scattering amplitude $f(\theta)$ can be also obtained simultaneously, but it is not needed in deriving the Sommerfeld factor. 

Without the potential term in Eq.~(\ref{eq:one_particle_schrodinger}) for all $r$, we get
\beq
R_{kl,free} (r) = 2k j_l (kr) \,,
\label{eq:R_l-free}
\eeq
where we have used the same normalization at $r \to \infty$ and the requirement that $R_{kl,free} (r) \propto r^l$ as $r \to 0$. 

Following the same procedure as in getting $A_l$, one can obtain 
\beq
A_{l,free} = \frac{1}{2 k} (2l+1) i^l  \,.
\label{eq:A_l-free}
\eeq
This is as expected, since there is no phase shift without a potential for all $r$. 
Indeed, Eqs. (\ref{eq:psi_free}), (\ref{eq:R_l-free}) and (\ref{eq:A_l-free}) give
\beq
\psi_{free} (\vec{r}) = \sum_{l=0}^\infty \frac{1}{2 k} (2l+1) i^l P_l (\cos\theta) 2k j_l (kr) = 
e^{ikz} \,.
\eeq 
From Eq.~(\ref{eq:R_l-free}), we have 
\beq
R_{kl,free} (r) \overset{r \to 0}{\longrightarrow} \frac{2 k^{l+1} r^l}{(2 l+1)!! } \,.
\eeq
Therefore Eq.~(\ref{eq:l-wave_Sommerfeld}) becomes
\beq
S_l = \Big[\frac{(2 l+1)!!}{2 k^{l+1}} \Big]^2 \lim_{r \to 0} \Big\lvert \frac{R_{kl} (r)}{r^l} \Big\rvert^2 = \Big[\frac{(2 l+1)!!}{2  k^{l+1} \, l!} \Big]^2 \lim_{r \to 0} \Big\lvert \frac{d^l R_{kl} (r)}{d r^l} \Big\rvert^2 \,,
\label{eq:l-wave_sommerfeld_factor}
\eeq
where in the last step we have used again the requirement that $R_{kl} (r) \propto r^l$ as $r \to 0$. 
\\

In the rest of this appendix, as an example, we derive Sommerfeld factors for a finite-range Coulomb potential, namely, 
\beq
V(r) = 
\begin{cases}
- \frac{\alpha}{r}, &r < r_0 \\
0, &r>r_0
\end{cases}
\label{eq:Coulomb}
\eeq
where $\alpha > 0$ for an attractive case, and $\alpha < 0$ for a repulsive case. 

For $r < r_0$, $R_{kl} (r)$ can be written as a linear combination of the Whittaker functions  
\beq
R_{kl} (r) = \frac{c_{l_{\rm inner1}}}{r} M_{- \frac{i}{\epsilon_v}, \, l + \frac{1}{2}}(2 i k r) 
+ \frac{c_{l_{\rm inner2}}}{r} W_{- \frac{i}{\epsilon_v}, \, l + \frac{1}{2}}(2 i k r) \,, \;\; (\text{for} \;\; r < r_0) \,,
\eeq
where $\epsilon_v \equiv \frac{v_{rel}}{\alpha}$. 
The condition Eq.~(\ref{eq:solution_condition_1}) requires that $c_{l_{\rm inner2}}= 0$, so that 
\beq
R_{kl} (r) = \frac{c_{l_{\rm inner1}}}{r} M_{- \frac{i}{\epsilon_v}, \, l + \frac{1}{2}}(2 i k r) \,, \;\; (\text{for} \;\; r < r_0) \,.  
\eeq
Using Eq.~(\ref{eq:outer_solution_normalized}) and the continuity of $R_{kl} (r)$ and $\frac{d R_{kl} (r)}{dr}$ at $r = r_0$, one can solve for $c_{l_{\rm inner1}}$ and $\delta_l$.
Using  
\beq
M_{- \frac{i}{\epsilon_v}, \, l + \frac{1}{2}}(2 i k r) \overset{r \to 0}{\longrightarrow} (2ikr)^{l+1} \,,
\eeq
we get from Eq.~(\ref{eq:l-wave_sommerfeld_factor}) the $l$-wave Sommerfeld factor 
\beq
S_{l_{Coulomb}} = 2^{2l} \, [(2l+1)!!]^2 \, |c_{l_{\rm inner1}}|^2 \,.
\eeq

For a general $l$, using the analytical expression of $|c_{l_{\rm inner1}}|^2$, we have checked that
\beq
S_{l_{Coulomb}} \overset{r_0 \to \infty}{\longrightarrow} e^{\pi/\epsilon_v} \frac{\pi/\epsilon_v}{\sinh(\pi/\epsilon_v) (l!)^2} \prod_{s=1}^{l} (s^2 + \epsilon_v^{-2}) \,,
\eeq
which is the result given in the literature for a Coulomb potential without truncation~\cite{Iengo:2009ni, Cassel:2009wt}. 
We have also checked that, as expected, 
\beq
S_{l_{Coulomb}} \overset{r_0 \to 0}{\longrightarrow} 1 \,.
\eeq

The $s$-wave Sommerfeld factor is 
\beq
S_{0_{Coulomb}} = \frac{\epsilon_v^2}{\Big\lvert 
\big[(1+ i \epsilon_v) f_1 -  
f_2 \big] 
\big[ (1+ i \epsilon_v ) f_1 - 
(1 + 2 \epsilon_v^2 \eta_0) f_2 \big]
\Big\rvert} \,,
\label{eq:s-wave_Coulomb}
\eeq
where $\eta_0 \equiv \alpha \mu r_0$,  
$f_1 \equiv F (\frac{i}{\epsilon_v},2,2 i \epsilon_v \eta_0)$ 
and 
$f_2 \equiv F (1 + \frac{i}{\epsilon_v},2,2 i \epsilon_v \eta_0)$. 
$F$ is the confluent hypergeometric function, given by
$F(a, b, z) = 1 + \frac{a}{b} \frac{z}{1!} + \frac{a(a+1)}{b(b+1)} \frac{z}{2!} + \cdots$.

\section{Results for a Hulth\'en potential}
\label{sec:appendix B}

Exchanges of some light but massive mediator between a pair of annihilating particles can give rise to a Yukawa-like potential. 
The main difference between Sommerfeld factors for a Coulomb and a Yukawa potential is that the latter feature resonances. 
The importance of these resonances in DM indirect searches and relic abundance calculations has been well-studied.
While the key physics that we want to convey has been illustrated in the main text by studying a Coulomb potential, we would like to investigate whether the resonance feature in Sommerfeld factors can bring more interesting results.  

The procedure to obtain Sommerfeld factors outlined in Appendix~\ref{sec:appendix A} can be applied to Yukawa potentials. 
However, the exact Sommerfeld factor for a Yukawa potential has to be obtained numerically. 
Fortunately, it is known that a Yukawa potential can be approximated by a Hulth\'en potential, for which an analytic Sommerfeld factor can be found~\cite{Cassel:2009wt}. 
Therefore, in this appendix we present results for a finite-range Hulth\'en potential,
\beq
V(r) = 
\begin{cases}
- \frac{\textstyle \alpha m_\phi^* e^{-m_\phi^* r}}{\textstyle 1- e^{-m_\phi^* r}}, &r < r_0 \\
0, &r>r_0
\end{cases}
\label{eq:Hulthen}
\eeq
where $\alpha > 0$ for an attractive case, and $\alpha < 0$ for a repulsive case. 
It was found~\cite{Cassel:2009wt} and confirmed~\cite{Feng:2010zp} that by relating $m_\phi^*$ with the mediator mass $m_\phi$ as $m_\phi^* = \frac{\pi^2 m_\phi}{6}$, the $s$-wave Sommerfeld factor for a Hulth\'en potential is an excellent approximation of the one for a Yukawa potential $- \alpha e^{-m_\phi r}/r$. 
  
Using the procedure described in Appendix~\ref{sec:appendix A}, we derive the analytic (though lengthy) $s$-wave Sommerfeld factor for this potential, $S_{0_{Hulthen}}$, which is the analogue of Eq.~(\ref{eq:s-wave_Coulomb}) for the finite-range Coulomb potential. 
We have checked that, as expected, 
\beq
S_{0_{Hulthen}} \overset{r_0 \to 0}{\longrightarrow} 1 \,.
\eeq
Also, 
\beq
S_{0_{Hulthen}} \overset{r_0 \to \infty}{\longrightarrow} 
\frac{2 \pi}{\epsilon_v} \frac{\sinh\big(\frac{2 \pi \epsilon_v}{y}\big)}{\cosh\big(\frac{2 \pi \epsilon_v}{y}\big) - \cos\big(2 \pi \sqrt{\frac{2}{y} - \frac{\epsilon_v^2}{y^2}}\big)}
 \equiv S_{r_0 \to \infty_{Hulthen}} \,,
\eeq
where $y \equiv m_\phi^*/(\mu \alpha)$, and this expression is the same as the one given in the literature for a Hulth\'en potential without truncation~\footnote{Note that the $\epsilon_v$ in Eq.~(6) of~\cite{Feng:2010zp} is $\epsilon_v/2$ in our notation.}. 

Substituting $S_{0_{Coulomb}}$ by $S_{0_{Hulthen}}$ in Eq.~(\ref{eq:sbar0Coulomb}), we obtain the $s$-wave rate-averaged Sommerfeld factor $\bar{S}_{0_{Hulthen}}$, which is to be compared with $S_{r_0 \to \infty_{Hulthen}}$. 
By further substituting $\bar{S}_{0_{Coulomb}}$ by $\bar{S}_{0_{Hulthen}}$ in Eq.~(\ref{eq:sigmavrel_thermal}), and $S_{r_0 \to \infty}$ by $S_{r_0 \to \infty_{Hulthen}}$ in Eq.~(\ref{eq:sigmavrel_thermal_infty}), we obtain the thermally averaged Sommerfeld factors $\langle \bar{S}_{0_{Hulthen}}\rangle$ and $\langle S_{r_0 \to \infty_{Hulthen}} \rangle$, respectively, for the Hulth\'en potential. 

Because for a massive mediator usually people are interested in the Sommerfeld enhancement and in particular the resonance behavior, in the following we show $s$-wave results for attractive Hulth\'en potentials and pay special attention to the largest resonance. 
To facilitate comparisons with the results of attractive Coulomb potentials shown in the main text, we use $\alpha = 0.1$ and the same two choices of $\xi$, namely, $0.01$ (black curves) and $0.1$ (purple curves). 

\begin{figure}
\begin{center}
\begin{tabular}{c c}
\hspace{-1.0cm}
\includegraphics[height=5.8cm]{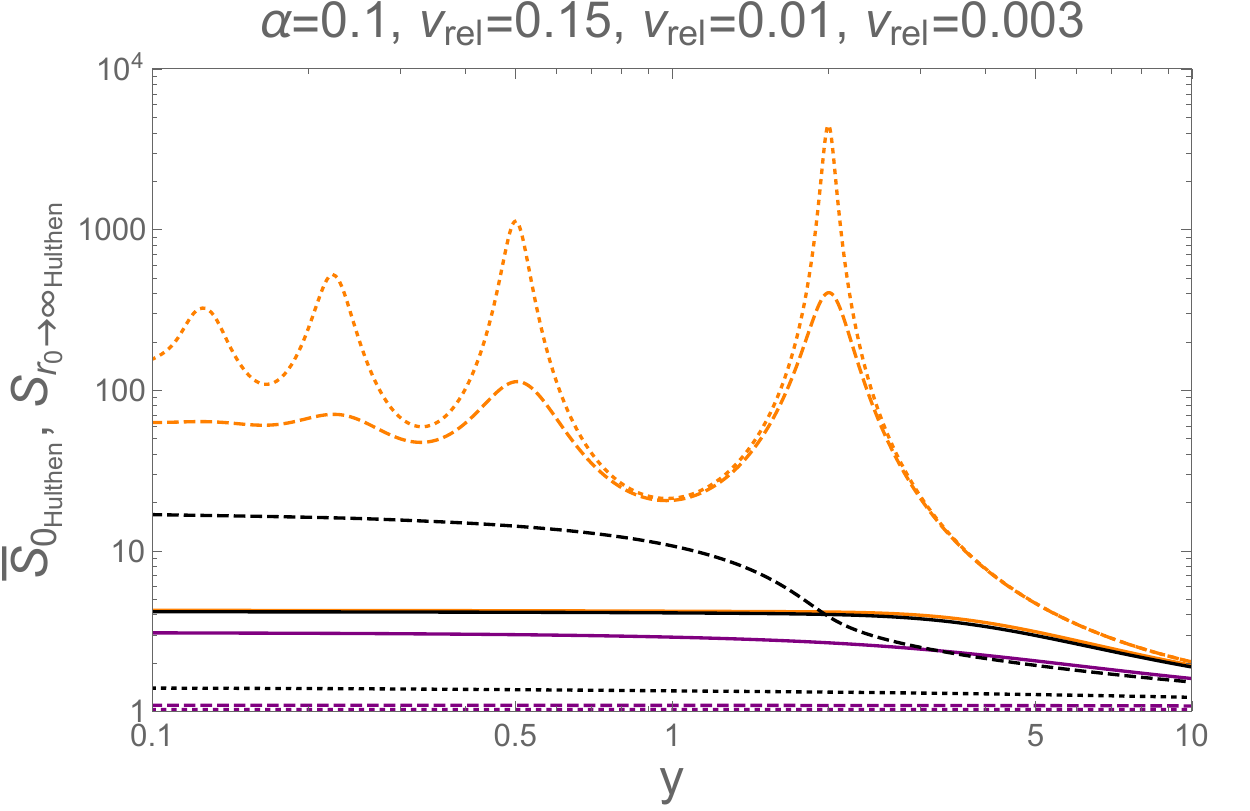} & 
\hspace{-0.2cm}
\includegraphics[height=5.8cm]{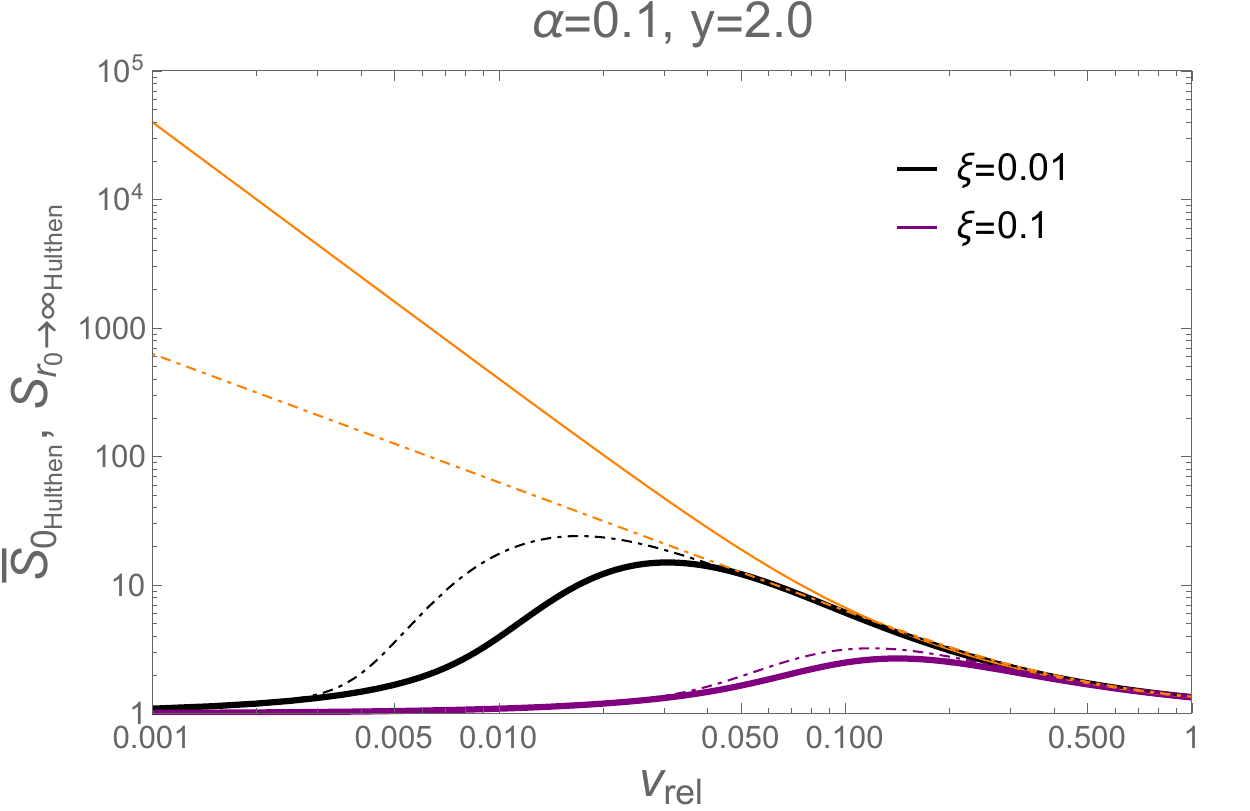} \\
\end{tabular}
\end{center}   
\caption{\label{fig:s-wave_attractive_Hulthen_bar}
\it
In the left panel, we show $\bar{S}_{0_{Hulthen}}$ (black and purple lines) and $S_{r_0 \to \infty_{Hulthen}}$ (orange lines) as functions of $y$, for three different $v_{rel}$. 
The solid, dashed and dotted lines are for $v_{rel} = 0.15$, $0.01$ and $0.003$, respectively.  
In the right panel, using solid lines we show $\bar{S}_{0_{Hulthen}}$ (black and purple) and $S_{r_0 \to \infty_{Hulthen}}$ (orange) as functions of $v_{rel}$, for $y = 2.0$. 
For comparison, the curves for an attractive Coulomb potential with the same $\alpha$ shown in the upper-left panel of Figure~\ref{fig:s-wave_Coulomb_bar} are replotted here using dot-dashed lines.
$\alpha = 0.1$ is used in both panels, and the black and purple lines are for $\xi = 0.01$ and $0.1$, respectively. 
}
\end{figure}

In the left panel of Figure~\ref{fig:s-wave_attractive_Hulthen_bar}, $\bar{S}_{0_{Hulthen}}$ and $S_{r_0 \to \infty_{Hulthen}}$ are plotted as functions of $y$ for three choices of $v_{rel}$, 0.15, 0.01 and 0.003. 
For $S_{r_0 \to \infty_{Hulthen}}$, it is known that the resonances are more prominent for smaller $\epsilon_v$, and indeed we can see the resonances clearly for $v_{rel} = 0.01$ and $0.003$. 
Similar to the attractive Coulomb case, for a given $v_{rel}$, $\bar{S}_{0_{Hulthen}}$ is smaller for larger $\xi$. 
Additionally, we see that the resonances are more suppressed for larger $\xi$. 

In the right panel of Figure~\ref{fig:s-wave_attractive_Hulthen_bar}, $\bar{S}_{0_{Hulthen}}$ and $S_{r_0 \to \infty_{Hulthen}}$ are plotted as functions of $v_{rel}$ for $y = 2.0$, which is the value close to the position of the highest peak in the left panel. 
The curves are similar to the ones in the upper-left panel of Figure~\ref{fig:s-wave_Coulomb_bar} for the $s$-wave result of an attractive Coulomb potential. 
To make the comparison easier, we replot the curves of the latter case using dot-dashed lines. 
We can see that on the small $v_{rel}$ side, without a truncation in the potentials, the Sommerfeld factor for a Hulth\'en potential is much larger than the one for a Coulomb potential. 
However, when a truncation is considered, the rate-averaged Sommerfeld factors for a Hulth\'en potential are smaller than the ones for a Coulomb potential. 
On the large $v_{rel}$ side, the resonance in the Hulth\'en case is almost invisible, and the lines with the same color merge. 
Other features of the curves for the Hulth\'en potential can be explained the same as the ones for the Coulomb potential, and we refer the reader to the discussions around the upper-left panel of Figure~\ref{fig:s-wave_Coulomb_bar} for details.

\begin{figure}
\begin{center}
\begin{tabular}{c c}
\hspace{-1.0cm}
\includegraphics[height=5.8cm]{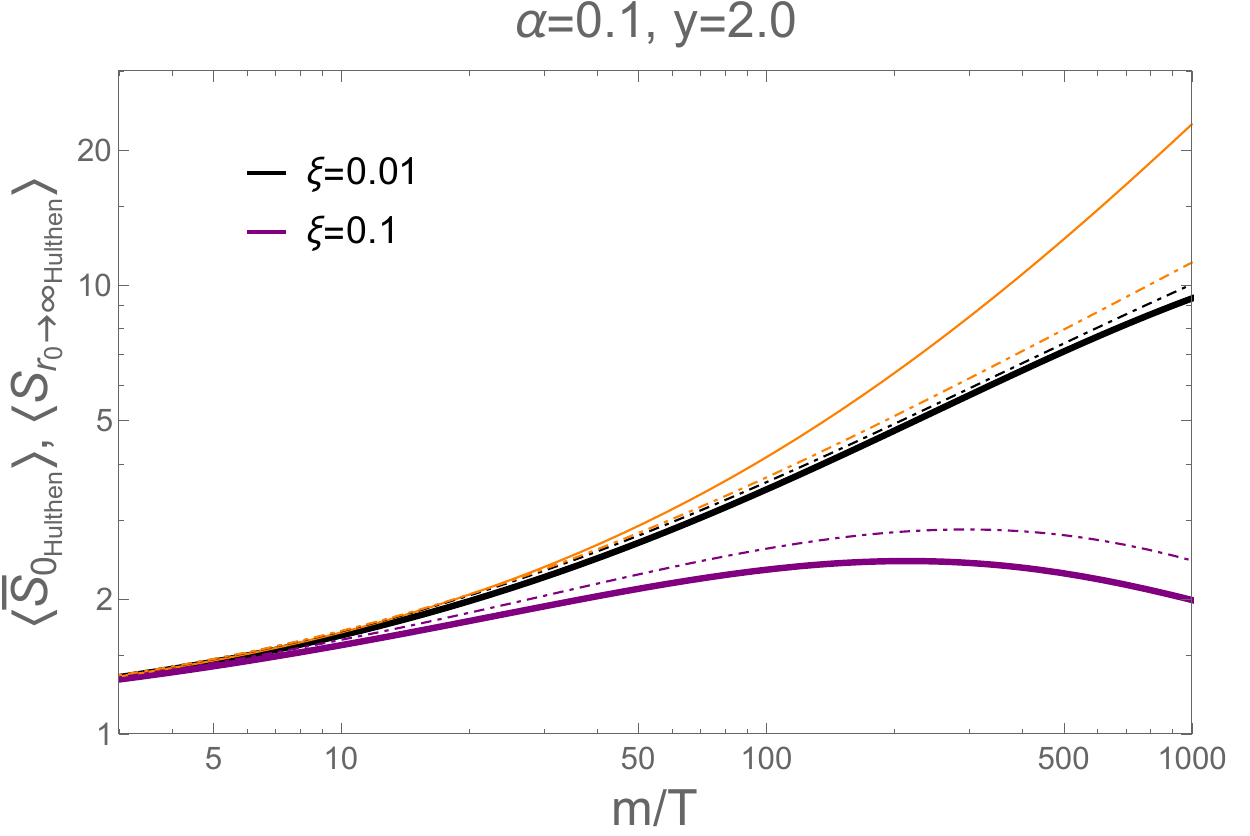} & 
\hspace{-0.2cm}
\includegraphics[height=5.8cm]{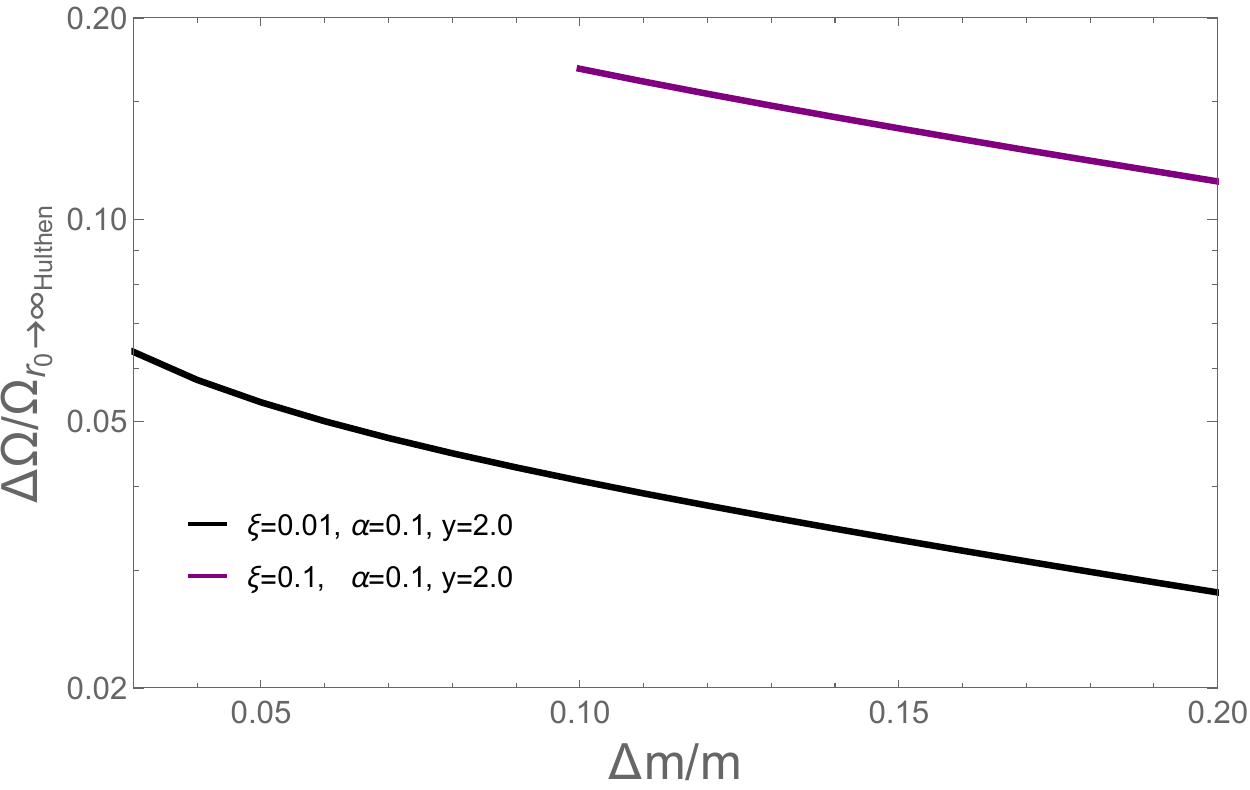} \\
\end{tabular}
\end{center}   
\caption{\label{fig:s-wave_attractive_Hulthen_moverT_and_Delta_ohsq}
\it
Left panel: $\langle \bar{S}_{0_{Hulthen}}\rangle$ (black and purple lines) and $\langle S_{r_0 \to \infty_{Hulthen}} \rangle$ (orange line) for an attractive Hulth\'en potential for a pair of unstable annihilating particles, as functions of the ratio of annihilating particle's mass to temperature.
Right panel: the relative change of the DM thermal relic abundance due to the modification of the Sommerfeld factor induced by coannihilators' decay, $\Delta \Omega/\Omega_{r_0 \to \infty_{Hulthen}}$, as a function of $\Delta m/m$, for $g_\chi = g_{\tilde{C}} = 2$, $m_{\rm DM} = 2 \times 10^3 \,\text{GeV}$ and $a_{\tilde{C} \tilde{C}} = 10^{-8} \, \text{GeV}^{-2}$.
In both panels, $\alpha = 0.1$ and $y = 2.0$ are used, and the black and purple lines are for $\xi = 0.01$ and $0.1$, respectively.
For comparison, in the left panel we replot using dot-dashed lines the curves for an attractive Coulomb potential with the same $\alpha$, shown by the solid lines in the left panel of Figure~\ref{fig:s-wave_Coulomb_bar_moverT}. 
}
\end{figure}

In the left panel of Figure~\ref{fig:s-wave_attractive_Hulthen_moverT_and_Delta_ohsq}, we plot $\langle \bar{S}_{0_{Hulthen}}\rangle$ and $\langle S_{r_0 \to \infty_{Hulthen}} \rangle$ as functions of $m/T$. 
These curves are similar to the solid ones shown in the left panel of Figure~\ref{fig:s-wave_Coulomb_bar_moverT}, which are replotted here using dot-dashed lines. 
The difference between the solid and the dot-dashed lines with the same color can be understood from the discussions for the right panel of Figure~\ref{fig:s-wave_attractive_Hulthen_bar}, by recalling that a larger $m/T$ corresponds to a smaller typical value of $v_{rel}$. 
Again, other features of the curves for the Hulth\'en potential can be explained the same as the ones for the Coulomb potential, and one can refer to the discussions for the left panel of Figure~\ref{fig:s-wave_Coulomb_bar_moverT}. 

In the right panel of Figure~\ref{fig:s-wave_attractive_Hulthen_moverT_and_Delta_ohsq}, we show $\Delta \Omega/\Omega_{r_0 \to \infty_{Hulthen}}$ as a function of $\Delta m/m$ for the same choice of parameters as in Figure~\ref{fig:Delta_ohsq}, namely, $g_\chi = g_{\tilde{C}} = 2$, $m_{\rm DM} = 2 \times 10^3 \,\text{GeV}$ and $a_{\tilde{C} \tilde{C}} = 10^{-8} \, \text{GeV}^{-2}$. 
$\Delta \Omega/\Omega_{r_0 \to \infty_{Hulthen}}$ is defined the same way as in Eq.~(\ref{eq:delta_Omega}), but for the Hulth\'en potential. 
Compared to the corresponding solid lines in Figure~\ref{fig:Delta_ohsq} for the attractive Coulomb case, it appears that the relative change of the DM thermal relic abundance is larger for the resonance in the Hulth\'en potential.

\section{A derivation of the Schr\"odinger equation for a pair of unstable annihilating particles}
\label{sec:appendix C}

In this appendix, we derive Eq.~(\ref{eq:one_particle_schrodinger}) in Appendix~\ref{sec:appendix A} for a pair of unstable annihilating particles. 
The Sommerfeld factors we use in this work are determined by solving for this equation. 
Although it is the common starting point in the literature to derive the Sommerfeld factors, some discussions are needed in the context when the annihilating particles can decay. 

The quantum mechanical approach in deriving Sommerfeld factors relies on the assumption that the distance scale for the long-range force and the one for annihilation are well separated. 
Generally, this assumption is met because the annihilation can be approximated to occur only when the two particles collide at $\vec{r} = 0$. 
However, when annihilating particles' decay is considered, one may wonder whether it is still possible to use this approach to derive Sommerfeld factors, given that now the decay is also a short-distance process. 

Let's start from the time-dependent Schr\"odinger equation for two particles, written in terms of the center-of-mass coordinate $\vec{R}$ and the relative coordinate $\vec{r}$, 
\beq
i \frac{\partial}{\partial t} \Psi (\vec{R}, \vec{r}, t) = \Big[-\frac{1}{2 M} \nabla_{\vec{R}}^2 -\frac{1}{2 \mu} \nabla_{\vec{r}}^2 + U(\vec{r}) - i \frac{\Gamma}{2} \Big] \Psi (\vec{R}, \vec{r}, t) \,,
\label{eq:time-dependent two particle schrodinger}
\eeq
where $\vec{R} = \frac{m_1 \vec{r}_1 + m_2 \vec{r}_2}{m_1 + m_2}$, $\vec{r} = \vec{r}_1 - \vec{r}_2$, $M = m_1 + m_2$ and $\mu = \frac{m_1 m_2}{m_1 + m_2}$. 
$\Gamma$ is the sum of decay rates of the two particles. 
$U(\vec{r})$ is the sum of $V(r)$ in Eq.~(\ref{eq:one_particle_schrodinger}) and the annihilation term which is proportional to $\delta^3(\vec{r})$~\footnote{An example of the whole potential term $(U(\vec{r}) - i \frac{\Gamma}{2})$ may be found in Eq.~(3.4) of~\cite{Matsumoto:2022ojl}.}. 

As mentioned in the discussion below Eq.~(\ref{eq:sbar0Coulomb}), 
we neglect the velocity dependence of $\Gamma$, since during and after freeze-out the typical $v_{rel}$ is non-relativistic.
Therefore, we treat $\Gamma$ as a constant.
Now taking advantage of this, we can solve the above equation using the method of separation of variables, 
\beq
\Psi (\vec{R}, \vec{r}, t) = \psi (\vec{R}, \vec{r}) e^{-i (E_T - i \frac{\Gamma}{2})t},
\label{eq:schrodinger time solution}
\eeq
where $E_T$ is the total energy of the two-particle system. 
$\psi (\vec{R}, \vec{r})$ satisfies
\beq
\Big[-\frac{1}{2 M} \nabla_{\vec{R}}^2 -\frac{1}{2 \mu} \nabla_{\vec{r}}^2 + U(\vec{r}) \Big] \psi (\vec{R}, \vec{r}) = E_T \psi (\vec{R}, \vec{r}) \,,
\label{eq:time-independent schrodinger}
\eeq
and it can be further separated as $\psi (\vec{R}, \vec{r}) = \psi_c (\vec{R}) \psi_r (\vec{r})$, in which $\psi_c (\vec{R})$ describes the motion of the mass center, 
\beq
- \frac{1}{2M} \nabla_{\vec{R}}^2 \psi_c (\vec{R}) = E_c \psi_c (\vec{R}) \,,
\eeq
and $\psi_r (\vec{r})$ describes the relative motion of the two-particle system, 
\beq
\Big[- \frac{1}{2\mu} \nabla_{\vec{r}}^2 + U(\vec{r}) \Big] \psi_r (\vec{r}) = E \psi_r (\vec{r}) \,.
\label{eq:relative schodinger equation}
\eeq
$E$ is the relative energy appearing in Eq.~(\ref{eq:one_particle_schrodinger}), and the center-of-mass energy $E_c = E_T - E$. 

From Eq.~(\ref{eq:relative schodinger equation}), the usual procedure in literature to derive Sommerfeld factors is then to neglect the annihilation term in $U(\vec{r})$, and solve for Eq.~(\ref{eq:one_particle_schrodinger})~\footnote{We note that although it is the common procedure in literature, using Eq.~(\ref{eq:one_particle_schrodinger}) rather than Eq.~(\ref{eq:relative schodinger equation}) in deriving Sommerfeld factors could in some cases lead to a too large Sommerfeld-enhanced annihilation cross section violating partial-wave unitary. This problem was addressed in~\cite{Blum:2016nrz}. Nevertheless, our main finding is that the decay can suppress the Sommerfeld enhancement factor, the unitarity violation is not of much concern, and therefore we follow the usual procedure for simplicity.}. 

Some discussion may be needed for the factor $e^{- \frac{\Gamma}{2} t}$ in Eq.~(\ref{eq:schrodinger time solution}). 
This factor leads to a reduction of the flux of the annihilating particle pair.
However, in the context of coannihilation, the decay (and inverse-decay) of the coannihilators is traditionally taken into account in the coupled set of Boltzmann equations for the DM and coannihilators. 
In this sense, decay and annihilation processes of the coannihilators are in fact already simultaneously considered, no matter whether there is a Sommerfeld factor for the annihilation of coannihilators. 
Also, for the purpose of calculating DM relic abundance, the coupled set of Boltzmann equations can be reduced to a single Boltzmann equation if the decay rate is much larger than the Hubble expansion rate. The single Boltzmann equation is obtained by adding each of the coupled Boltzmann equations. 
In this way the decay (and inverse-decay) terms cancel.
Therefore, one usually does not have to worry about this factor.

\bibliographystyle{bibstyle_jhep}
\bibliography{ref_CL2}
\end{document}